\documentclass[onecolumn,noshowpacs,nofootinbib,aps,prl,superscriptaddress]{revtex4}

\usepackage{graphicx}
\usepackage{dcolumn}
\usepackage{amsmath}
\usepackage{epsfig}

\input{pubboard/babarsym.tex}

\def\phm{\phantom{-}}
\def\ctwob{\ensuremath{\cos\! 2 \beta   }\xspace}

\def\ncand{26334}
\def\nsig{\ensuremath{5098 \pm 102}}
\renewcommand{\eqref}[1]{Eq.~(\ref{eq:#1})}

\newcommand{\splots}   {\mbox{$_s{\cal P}lots$}\xspace}
\newcommand{\splot}    {\mbox{$_s{\cal P}lot$}\xspace}

\newcommand{\onreslumi}  {\mbox{429\invfb}}
\newcommand{\offreslumi} {\mbox{45\invfb}}
\newcommand{\bbpairs}    {\mbox{$470.9\pm2.8$~million}}

\newcommand{\Lzero}      {\mbox{$L_0$}}
\newcommand{\Ltwo}       {\mbox{$L_2$}}
\newcommand{\LtwoOverLzero} {\mbox{$\Ltwo/\Lzero$}}

\newcommand{\abscosbmom} {\mbox{$\abs{\cos{\theta_{\B \, {\rm mom}}}}$}}

\newcommand{\abscosbthr} {\mbox{$\abs{\cos{\theta_{\B \, {\rm thr}}}}$}}

\newcommand{\mpr}        {\mbox{$M$}}
\newcommand{\thpr}       {\mbox{$\Theta$}}

\newcommand{\mAC}         {\mbox{$m_{+}$}}
\newcommand{\mBC}         {\mbox{$m_{-}$}}

\newcommand{\mACSq}       {\mbox{$m^2_{+}$}}
\newcommand{\mBCSq}       {\mbox{$m^2_{-}$}}

\newcommand{\mACSqPr}     {\mbox{$\tilde{m}^2_{+}$}}
\newcommand{\mBCSqPr}     {\mbox{$\tilde{m}^2_{-}$}}

\newcommand{\cBC}         {\mbox{$\cos\theta_{-}$}}

\newcommand{\Dstarv}             {\mbox{$\Dstar_{v}(2010)$}}

\newcommand{\Dstarvm}            {\mbox{$\Dstar_{v}(2010)^-$}}
\newcommand{\Dstarvmpip}          {\mbox{$\Dstarvm \pip$}}

\newcommand{\DstarI}             {\mbox{$\Dstar_{0}(2400)$}}

\newcommand{\DstarIm}            {\mbox{$\Dstar_{0}(2400)^-$}}
\newcommand{\DstarImpip}          {\mbox{$\DstarIm \pip$}}
\newcommand{\BztoDstarImpip}      {\mbox{$\Bz \to \DstarImpip$}}

\newcommand{\DstarII}            {\mbox{$\Dstar_{2}(2460)$}}
\newcommand{\DstarIIp}           {\mbox{$\Dstar_{2}(2460)^+$}}
\newcommand{\DstarIIm}           {\mbox{$\Dstar_{2}(2460)^-$}}
\newcommand{\DstarIImpip}         {\mbox{$\DstarIIm \pip$}}
\newcommand{\BztoDstarIImpip}     {\mbox{$\Bz \to \DstarIImpip$}}

\def\Kpi   {\ensuremath{K^+\pi^-}\xspace}

\newcommand{\rhoz}               {\mbox{$\rho^0$}}
\newcommand{\rhop}               {\mbox{$\rho^+$}}

\newcommand{\rhoI}               {\mbox{$\rho(770)$}}
\newcommand{\rhoIz}              {\mbox{$\rho(770)^0$}}

\newcommand{\rhoIzDzb}           {\mbox{$\rhoIz \Dzb$}}

\newcommand{\omegaI}              {\mbox{$\omega(782)$}}

\newcommand{\fz}                 {\mbox{$f_0$}}
\newcommand{\fI}                 {\mbox{$\fz(980)$}}

\newcommand{\rhoII}              {\mbox{$\rho(1450)$}}

\newcommand{\fII}                {\mbox{$f_2(1270)$}}

\newcommand{\fIIDzb}             {\mbox{$\fII \Dzb$}}

\newcommand{\fIII}               {\mbox{$f_0(1370)$}}

\newcommand{\D}                  {\mbox{$D$}}

\newcommand{\Dzbpipi}            {\mbox{$\Dzb \pipi$}}
\newcommand{\BztoDzbpipi}        {\mbox{$\Bz \to \Dzbpipi$}}

\newcommand{\DzbtoKpi}           {\mbox{$\Dzb \to \Kp\pim$}}

\newcommand{\abs}[1]{\ensuremath{\left|#1\right|}}

\newcommand{\fcc}[1]{\multicolumn{4}{c}{#1}}
\newcommand{\fvcc}[1]{\multicolumn{5}{c}{#1}}

\graphicspath{{Figures/}}

\newcommand{\BABARPubYear}    {10}
\newcommand{\BABARPubNumber}  {004}
\newcommand{\SLACPubNumber} {14203}

\begin{document}

\preprint{\babar-PUB-\BABARPubYear/\BABARPubNumber} 
\preprint{SLAC-PUB-\SLACPubNumber} 

\begin{flushleft}
\babar-CONF-\BABARPubYear/\BABARPubNumber\\
SLAC-PUB-\SLACPubNumber\\
\end{flushleft}

\title{
  \large \bf \boldmath Dalitz-plot Analysis of $\Bz\to\Dzb\pipi$
}

%
\author{P.~del~Amo~Sanchez}
\author{J.~P.~Lees}
\author{V.~Poireau}
\author{E.~Prencipe}
\author{V.~Tisserand}
\affiliation{Laboratoire d'Annecy-le-Vieux de Physique des Particules (LAPP), Universit\'e de Savoie, CNRS/IN2P3,  F-74941 Annecy-Le-Vieux, France}
\author{J.~Garra~Tico}
\author{E.~Grauges}
\affiliation{Universitat de Barcelona, Facultat de Fisica, Departament ECM, E-08028 Barcelona, Spain }
\author{M.~Martinelli$^{ab}$}
\author{A.~Palano$^{ab}$ }
\author{M.~Pappagallo$^{ab}$ }
\affiliation{INFN Sezione di Bari$^{a}$; Dipartimento di Fisica, Universit\`a di Bari$^{b}$, I-70126 Bari, Italy }
% ...
\collaboration{The \babar\ Collaboration}
\noaffiliation

\date{\today}

\begin{abstract}
  We report preliminary results from a study of the decay $\Bz\to\Dzb\pipi$
  using a data sample of \bbpairs\ \BB\ events collected with the
  \babar\ detector at the \FourS\ resonance.
  Using the Dalitz-plot analysis technique, we find contributions from 
  the intermediate resonances \DstarIIm, \DstarIm, \rhoIz\ and \fII\ as well 
  as a \pipi\ S-wave term, a $\Dzb\pim$ nonresonant S-wave term and a virtual 
  $\Dstar(2010)^-$ amplitude.
  We measure the branching fractions of the contributing decays.
\end{abstract}

\maketitle

\section{Introduction}

The study of the Dalitz plot~\cite{Dalitz:1953cp} of $\Bz\to\Dzb\pipi$ decays
is motivated by several factors.
The branching fractions of $\B \to D^{**}$ transitions\footnote{
  $D^{**}$ mesons are P-wave excitations of states containing one charmed and
  one light ($u$, $d$) quark.
} are of interest to help address a conflict between theoretical
predictions~\cite{Jugeau:2005yr} and experimental
results~\cite{:2007rb,Aubert:2008ea} in semileptonic $\B\to D^{**}l\nu$ decays.
The $\Dzb\pipi$ final state allows relatively clean studies of the $J^P = 0^+$
and $2^+$ $D^{**}$ states, since the $1^+$ mesons cannot decay to $D\pi$.
Measurements of these decays test theoretical models including 
quark models~\cite{Morenas:1997nk}, 
QCD sum rules~\cite{Dai:1998ca,Uraltsev:2000ce,LeYaouanc:2001nk} and 
lattice QCD~\cite{Blossier:2009vy}. 
Similarly, measurement of the branching fraction of the $\Bz\to\Dzb\rhoz$
decay will help to test the dynamics of ``color-suppression'' in \B\ decays
(related to the fact that the color quantum numbers of the quarks produced from
the virtual $W$ boson must match that of the spectator quark in order for a
\rhoz\ meson to be
formed)~\cite{Bauer:1986bm,Neubert:2001sj,Chua:2001br,Mantry:2003uz}.
Moreover, using isospin symmetry to relate the decay amplitudes of
$\Bz\to\Dzb\rhoz$, $\Bz\to\Dm\rhop$ and $\Bp\to\Dzb\rhop$, it is possible to
study effects of final state interactions in these
decays~\cite{Rosner:1999zm,Neubert:2001sj}.

Another motivation is that the $\Bz\to\Dzb\rhoz$ decay can be used to measure
$\stwob$, where $\beta$ is the CKM unitarity triangle
angle~\cite{Cabibbo:1963yz,Kobayashi:1973fv}, if the \Dzb\ meson is
reconstructed in a \CP\ eigenstate. 
The measurement of this angle in the $\bar{b} \to \bar{c} u \bar{d}$
quark-level transition is theoretically cleaner than the commonly used 
$\bar{b} \to \bar{c} c \bar{s}$ decays (such as
$\Bz\to\jpsi\KS$)~\cite{Fleischer:2003ai,Fleischer:2003aj} 
and comparisons of the values measured in different quark-level transitions
can be used to search for the influence of physics beyond the Standard
Model~\cite{Grossman:1996ke}.
The time-dependent analysis of the $\Bz\to\Dzb\pipi$ Dalitz plot not only 
allows a proper handling of effects due to interference between broad
resonances, but also enables an improved measurement of $\beta$ since terms
proportional to $\ctwob$ as well as \stwob\ can be
measured~\cite{Charles:1998vf,Latham:2008zs}. 
For such an analysis, it is necessary to have a good understanding of the
population of the $\Bz\to\Dzb\pipi$ Dalitz plot.
This can be best studied in the \DzbtoKpi\ decay, which is the subject of this
study.

The $\Bz\to\Dzb\pipi$ decay has been previously studied by
Belle~\cite{Kuzmin:2006mw} and the related $\Bp\to\Dm\pip\pip$ decay has been
studied by both \babar~\cite{Aubert:2009wg} and Belle~\cite{Abe:2003zm}.
In this paper we present preliminary results from the first study of the
$\Bz\to\Dzb\pipi$ decay by \babar. 
The data used in the analysis, collected with the
\babar\ detector~\cite{Aubert:2001tu} at the \pep2\ asymmetric energy 
\epem\ collider at SLAC, consist of an integrated luminosity of
\onreslumi\ recorded at the \FourS\ resonance (``on-peak'') and
\offreslumi\ collected 40\,\mev\ below the resonance (``off-peak''). 
The on-peak data sample contains the whole \babar\ dataset of
\bbpairs\ \BB\ events.

\section{Selection}


We reconstruct $\Bz\to\Dzb\pipi$ candidates (the inclusion of charge conjugate
reactions is implied throughout this paper) by combining a \Dzb\ candidate
with two oppositely charged pion candidates.   
The charged pion candidates are required to satisfy particle
identification requirements that have efficiency above $97\,\%$ and kaon
misidentification probability below $20\,\%$.
We reconstruct $\Dzb$ mesons in the decay channel $\Kp\pim$.
For the $\Dzb$ daughters, the charged kaon candidates are required to satisfy
particle identification requirements that have efficiency above $97\,\%$ and
pion misidentification probability below $15\,\%$, while the charged pion
candidates are required to pass slightly looser criteria
than those for the bachelor pions.
The \D\ candidates are required to have an invariant mass within $15\mevcc$
of the nominal \Dzb\ mass~\cite{Amsler:2008zz}; this requirement is $85\,\%$
efficient for signal Monte Carlo (MC) events.


The $\Dzb$ candidate and the two bachelor pion candidates are required to
originate from a common vertex.
Signal events are distinguished from background 
using two almost uncorrelated
kinematic variables: the difference \DeltaE\ between the CM energy of the 
\B\ candidate and $\sqrt{s}/2$, and the beam-energy-substituted mass
$\mes=\sqrt{s/4-{\bf p}^2_\B}$, where $\sqrt{s}$ is the total CM energy and
${\bf p}_\B$ is the momentum of the candidate \B\ meson in the CM frame.
We apply preselection criteria of $-0.075 \gev < \DeltaE < 0.075 \gev$ 
and $5.272 \gevcc < \mes < 5.286 \gevcc$; these requirements are $86\,\%$
efficient for signal MC events.
We make further use of these kinematic variables to discriminate signal from
background in the fit described below. 
We exclude candidates consistent with the abundant $\Bz\to\Dstar(2010)^-\pip$
decay by rejecting events which contain a candidate with $\Dzb\pim$ invariant
mass below $2.02\gevcc$ (to maintain the symmetry of the Dalitz plot, we also
remove the region with $\Dzb\pip$ invariant mass below the same value). 
These events are used as a control sample to monitor differences between data
and MC.


To suppress the background contribution from continuum
$\epem\to\qqbar\ (\q=u,d,s,c)$ events, we construct a neural network (NN)
discriminant that combines four variables commonly used to
separate jet-like \qqbar\ events from the more spherical \BB\ events. 
These are: the $0^{\rm th}$ order momentum-weighted monomial moment,\footnote{
  The momentum-weighted monomial moments are defined 
  $L_j = \sum_i p_i \left| \cos \theta_i \right|^j$,
  where $\theta_j$ is the angle of the track or neutral cluster $i$ with
  respect to the signal \B\ thrust axis, $p_i$ is its momentum,
  and the sum excludes the daughters of the $B$ candidate.
} \Lzero;
the ratio of the $2^{\rm nd}$ order momentum-weighted monomial (\Ltwo) to that
of $0^{\rm th}$ order (\Lzero), \LtwoOverLzero; 
the absolute value of the cosine of the angle between the \B\ direction and
the beam ($z$) axis, \abscosbmom;
and the absolute value of the cosine of the angle between the \B\ thrust axis
and the beam ($z$) axis, \abscosbthr.
All these variables are evaluated in the \epem\ center-of-mass frame.
We apply a requirement on the NN output that retains approximately $88\,\%$ of
the signal and rejects $\sim 52\,\%$ of the continuum background.  
Most of the remaining background originates from \B\ decays, and is discussed
below.


After applying all selection criteria, we retain \ncand\ events with candidate
$\Bz\to\Dzb\pipi$ decays.
Around 20\% of these events have multiple candidates.
When an event has multiple candidates we retain the candidate
with the best geometrical \B-vertex probability.

\begin{figure}[!htb]
  \begin{tabular}{cc}
   \includegraphics[width=0.49\textwidth]{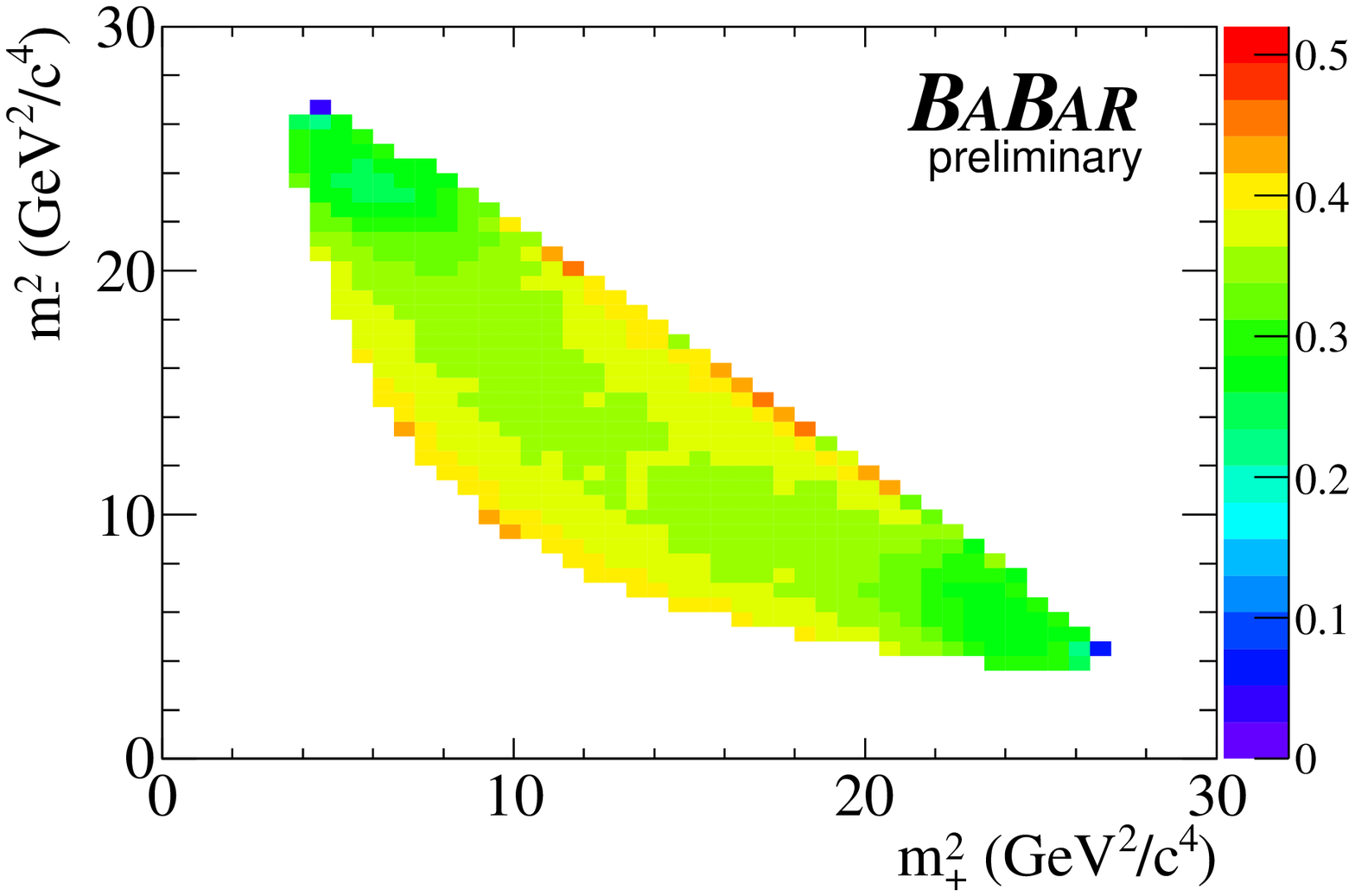} &
   \includegraphics[width=0.49\textwidth]{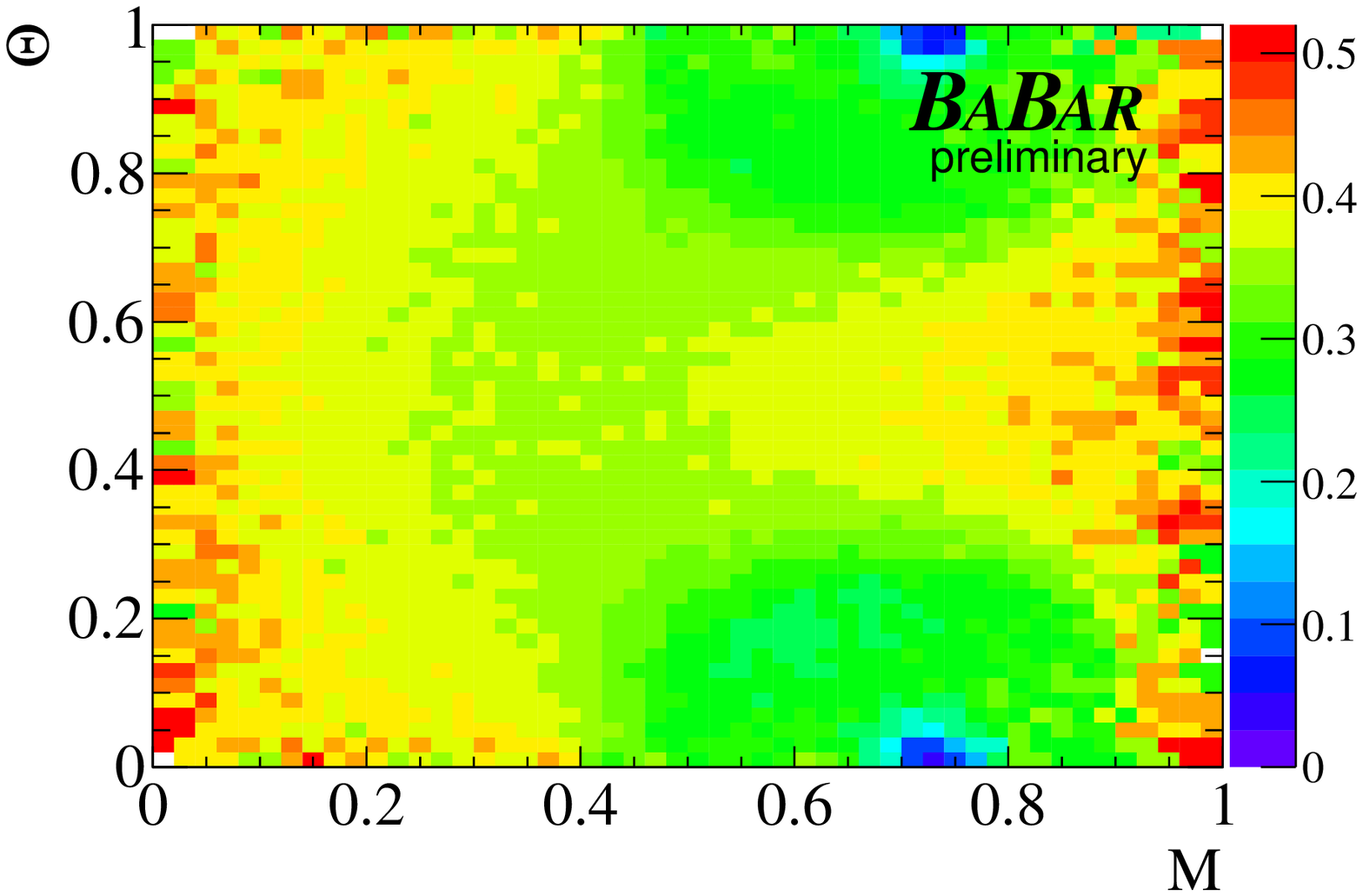}
  \end{tabular}
  \caption{
    Variation of the signal reconstruction efficiency over the phase-space.
    In these plots the $\Dstar(2010)^\pm$ veto is not applied.
  }
  \label{fig:eff-dp}
\end{figure}

The efficiency for signal events to pass all the selection criteria is
determined as a function of position in the Dalitz plot (DP).
Using a Monte Carlo (MC) simulation in which events uniformly populate the
phase-space, we obtain an average efficiency of approximately $35\,\%$.
The efficiency is shown as a function of phase-space in \figref{eff-dp}, both
in terms of the conventional DP (for which we choose axes 
$\mACSq = m^2_{\Dzb\pip}$ and $\mBCSq = m^2_{\Dzb\pim}$), 
and in terms of the ``square Dalitz plot'' (SDP).  
The latter is described by the variables \mpr\ and \thpr, 
\begin{equation}
  \label{eq:sqdp-vars}
  \mpr \equiv \frac{1}{\pi}
  \arccos\left(
    2\frac{
      m_{\pipi} - m^{\rm min}_{\pipi}
    }{
      m^{\rm max}_{\pipi} - m^{\rm min}_{\pipi}
    } - 1 
  \right)
  \ \ \ {\rm and} \ \ \ 
  \thpr \equiv \frac{1}{\pi}\theta_{\pipi}\, ,
\end{equation}
where $m_{\pipi}$ is the invariant mass of the two pions,
$m^{\rm max}_{\pipi} = m_{\Bz} - m_{\Dzb}$ and $m^{\rm min}_{\pipi} =
2 m_{\pi}$ are the kinematical limits of $m_{\pipi}$,
and $\theta_{\pipi}$ is the angle between the $\Dzb$ and the $\pip$
in the \pipi\ rest frame.
While the conventional DP representation provides a useful visual
representation of the physics of the signal decay, the SDP allows closer
scrutiny of the most densely populated regions of the phase-space, and hence
is appropriate for studies of background distributions, for example. 

\section{Backgrounds}

In addition to the background from continuum processes, we expect backgrounds
from other \BB\ decays.
These are studied using large MC samples in which the \B\ mesons decay
generically according to our current knowledge of their branching fractions.
We classify backgrounds from \BB\ decays in six categories based on their
\DeltaE\ and \mes\ distributions as determined from MC samples.
The different \BB\ background categories also have different DP distributions.

\tabref{backgrounds} lists the expected number of events and the dominant
contributing mode for each category. 
Background categories 1 and 2 have four track final states, and peak in both
\DeltaE\ and \mes\ -- category 1 has signal-like peaks in both, while category
2 has a \DeltaE\ peak shifted to positive values due to pion$\to$kaon
misidentification.
The decay modes that contribute to these categories do not contain real $D$
mesons, with the exception of $\Dzb\KS$, which contributes to category 1.
Background categories 3--6, which are dominant, do contain real $D$ mesons.
Category 3 peaks in \mes\ and has a \DeltaE\ distribution that is shifted to
negative values due to kaon$\to$pion misidentification.
Categories 4--6 do not peak strongly in either \DeltaE\ or \mes.
Category 4 has a broad \mes\ distribution and a slight peak in \DeltaE, and
includes background from $\Dstar(2010)^-\pip$ events that escape the veto due
to misreconstruction. 
Category 5 has a broad \mes\ distribution (similar to category 4) and an
approximately linear \DeltaE\ shape.
Category 6 has combinatorial distributions for both \mes\ and \DeltaE.
The continuum background shape is combinatorial and does not peak strongly in 
either \DeltaE\ or \mes.
A summary of the backgrounds in given in \tabref{backgrounds}.

\begin{table}
  \caption{
    Summary of backgrounds.  For each category the dominant contributing mode
    and the expected number of events after all selection requirements are
    applied to the data are given.
  }
  \label{tab:backgrounds}
  \begin{tabular}{l@{\hspace{5mm}}c@{\hspace{5mm}}c}
    \hline
    Category & Dominant contribution & Total \# Expected \\
    \hline
    \BB\ 1 & $\jpsi\Kpi$ & $444 \pm 24$ \\
    \BB\ 2 & $a_1^\pm\pimp$ & $32 \pm 7$ \\
    \BB\ 3 & $\Dz\Kpi$ & $240 \pm 18$ \\
    \BB\ 4 & $\Dzb\rhop$ & $7415 \pm 101$ \\
    \BB\ 5 & $\Dstarzb\pip$ & $1475 \pm 44$ \\
    \BB\ 6 & Combinatoric & $7336 \pm 99$ \\
    \hline
    \multicolumn{2}{l}{\qqbar} & $5352 \pm 226$  \\
    \hline
  \end{tabular}
\end{table}

\section{Maximum Likelihood Fit}

We perform an extended unbinned maximum likelihood fit using the variables
\DeltaE, \mes\ and the DP co-ordinates in order to determine the
signal yield and the properties of the Dalitz plot.
The complete likelihood function is given by:
\begin{equation}
  {\cal L} = \exp\left(-\sum_k N_k\right) 
  \prod_i^{N_e}
  \Bigg[
  \sum_k N_k {\cal P}_k(\mACSq^i,\mBCSq^i,\mes^i,\DeltaE^i)
  \Bigg] \,,
\end{equation}
where $N_k$ is the event yield for species $k$, the index $i$ runs over the
$N_e$ events in the data sample and ${\cal P}_k$ is the probability density
function (PDF) for species $k$, which consists of a product of the DP, 
\mes\ and \DeltaE\ PDFs.  
The different species $k$ are signal, \qqbar\ background and 
six \BB\ background categories.
The function $-\ln{\cal L}$ is minimized to obtain the preferred values of
the free parameters of the fit.

For each of the \BB\ background categories, the \DeltaE, \mes\ and DP PDFs are
described with histograms obtained using MC.
For \qqbar\ background, the \DeltaE\ and \mes\ PDFs are a $1^{\rm st}$-order
polynomial and an ARGUS function~\cite{Albrecht:1990cs}, respectively.
The parameters of the ARGUS function are fixed to values determined using
off-peak data, while the slope of the \qqbar\ \DeltaE\ PDF is a free parameter
of the fit. 
The continuum background DP PDF is modelled with a histogram obtained
from data in a sideband region of \mes, after subtraction of the (MC-based)
expected contribution from \BB\ decays in this region.
We have verified the consistency of our background PDFs in off-peak data, in
background MC samples, and in on-peak data sidebands.
All histograms used in the fit are in the square Dalitz plot format.

\begin{figure}[!htb]
  \begin{tabular}{cc}
   \includegraphics[width=0.49\textwidth]{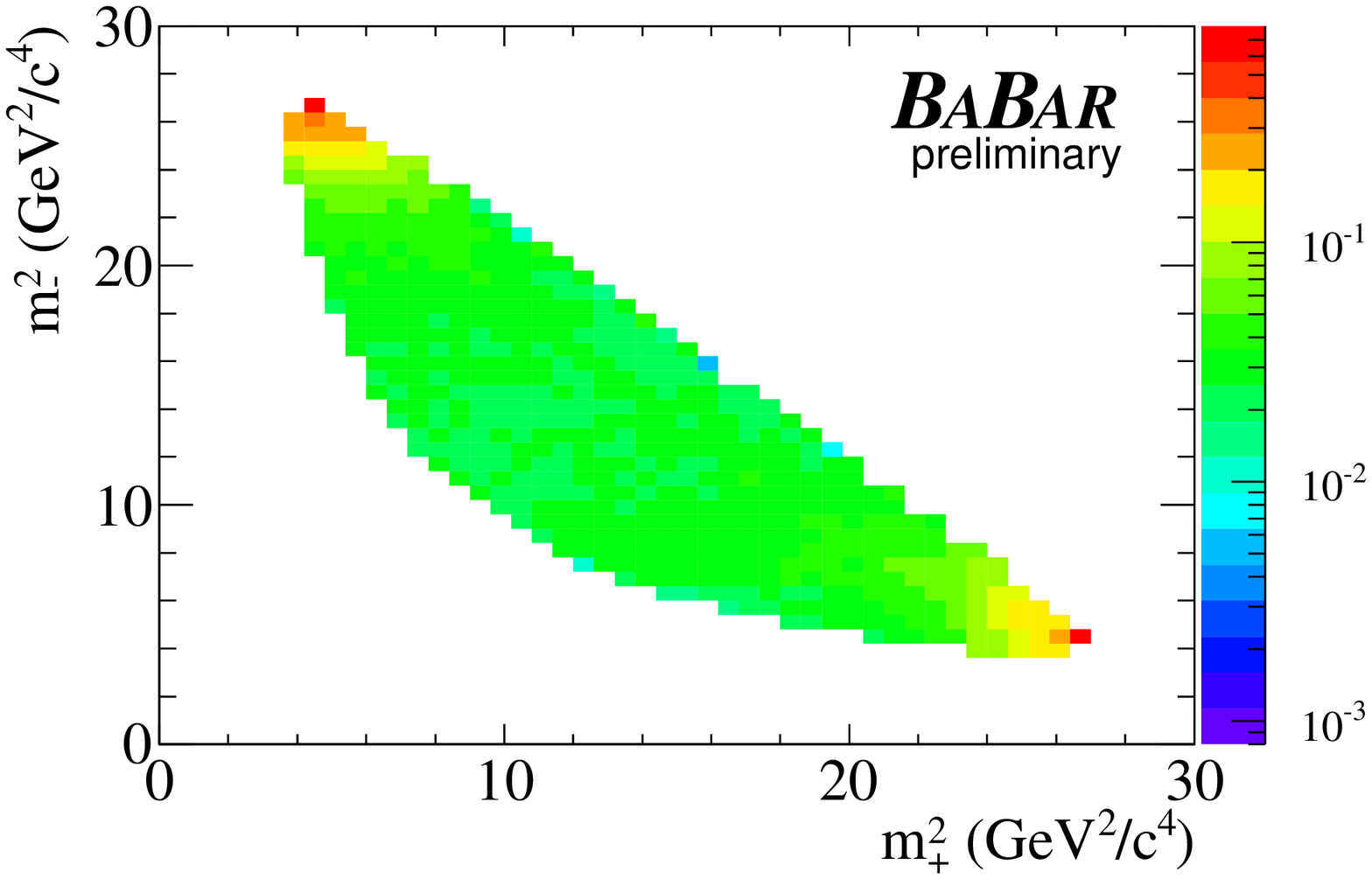} &
   \includegraphics[width=0.49\textwidth]{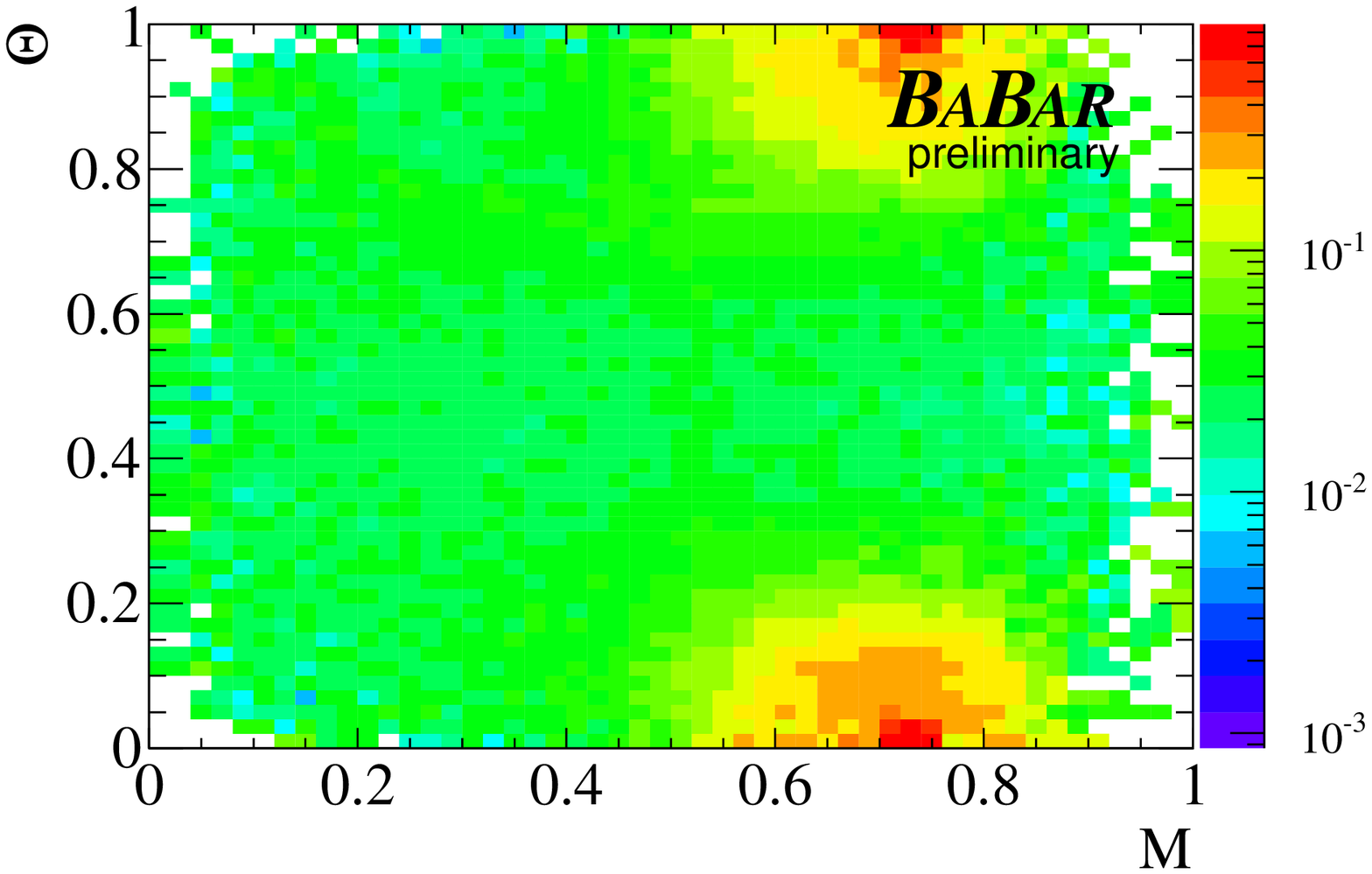}
  \end{tabular}
  \caption{
    Fraction of misreconstructed events as a function of the phase-space.
    In these plots the $\Dstar(2010)^\pm$ veto is not applied.
  }
  \label{fig:fscf}
\end{figure}

The signal component is composed of two parts which are distinguished by
whether or not the kinematics of the daughter particles are well
reconstructed.
We refer to the well reconstructed events as ``correctly reconstructed'' (CR)
and the misreconstructed events as ``self-cross-feed'' (SCF).
The fraction of SCF events as a function of DP position
$f_{\rm SCF}(\mACSq,\mBCSq)$ is determined from MC, and is shown in
\figref{fscf}.  Its value is typically below $10\,\%$ but is larger in the
corners of the Dalitz plot where one of the pions has low momentum. 

Both CR and SCF events have the same underlying physics PDF, but due to
misreconstruction SCF events have reconstructed DP positions that differ from
their true values.
This smearing is implemented by convoluting the PDF with a resolution function 
$R_{\rm SCF}(\mACSq,\mBCSq;\mACSqPr,\mBCSqPr)$
that gives the probability that an event with true DP position
$(\mACSqPr,\mBCSqPr)$ is reconstructed at $(\mACSq,\mBCSq)$,
and is described by a histogram in the square Dalitz plot co-ordinates that is
itself a function of position in the phase-space.  
For correctly reconstructed events, DP resolution effects are negligible. 
The signal Dalitz plot PDF is thus written as
\begin{eqnarray}
  {\cal P}_{\rm sig}(\mACSq,\mBCSq) & = &  
  \frac{1}{\cal N}
  \Bigg\{
      {\cal P}^{\rm phys}(\mACSq,\mBCSq) 
      \epsilon(\mACSq,\mBCSq)
      (1 - f_{\rm SCF}(\mACSq,\mBCSq))
      \, + \\
      & & 
      \hspace{-10mm}
      \int_{\rm DP} \, 
      \Big[
        {\cal P}^{\rm phys}(\mACSqPr,\mBCSqPr)
        \epsilon(\mACSqPr,\mBCSqPr) 
        f_{\rm SCF}(\mACSqPr,\mBCSqPr)
        R_{\rm SCF}(\mACSq,\mBCSq;\mACSqPr,\mBCSqPr) 
        \Big]
      d(\mACSqPr) d(\mBCSqPr)
      \Bigg\} \, , \nonumber
\end{eqnarray}
where ${\cal P}^{\rm phys}(\mACSq,\mBCSq)$ is the underlying physics PDF
(discussed below), 
$\epsilon(\mACSq,\mBCSq)$ is the efficiency (\figref{eff-dp}), and
$f_{\rm SCF}(\mACSq,\mBCSq)$ is the SCF fraction (\figref{fscf}).
The integral is over the Dalitz plot.
The normalization factor ${\cal N}$ ensures that 
${\cal P}_{\rm sig}(\mACSq,\mBCSq)$ gives unity when integrated over the
phase-space.

The CR and SCF signal events have different distributions in \DeltaE\ and \mes.
For \mes, both CR and SCF PDFs are described by double Gaussian functions
where the widths of the two Gaussians are constrained to be the same.   
For \DeltaE, the CR PDF is again a double Gaussian function (in which the two
Gaussians have different widths) while the SCF PDF is represented by an
asymmetric Gaussian with power-law tails.
The two Gaussian widths of the \DeltaE\ PDF for the CR component are given by
linear functions of 
$\left(m^{\rm min}_{D\pi}\right)^2 = {\rm min}(\mACSq,\mBCSq)$ 
to account for the momentum dependence of the resolution across the DP.
All SCF PDF parameters are fixed to values determined from MC,
while CR PDF parameters are floated in the fit where possible.
The CR PDF parameters that cannot be determined from the fit are determined
from MC.
Data/MC correction factors determined from the $\Dstar(2010)^-\pip$ control
sample are applied to all such parameters, except for the slopes of the
dependence of the \DeltaE\ widths on $\left(m^{\rm min}_{D\pi}\right)^2$.

We determine a nominal signal DP model using information from previous
studies of $\Bz\to\Dzb\pipi$~\cite{Kuzmin:2006mw} and
$\Bp\to\Dm\pip\pip$~\cite{Aubert:2009wg,Abe:2003zm}, 
and the change in the fit likelihood value observed when omitting or adding
resonances. 
We use the isobar model~\cite{Fleming:1964zz,Morgan:1968zz,Herndon:1973yn},
which models the total amplitude as resulting from a sum of amplitudes from
the individual decay channels:
\begin{equation}
  {\cal P}^{\rm phys}(\mACSq,\mBCSq) = 
  \left| {\cal A}(\mACSq,\mBCSq) \right|^2
  \hspace{5mm}
  {\rm where}
  \hspace{5mm}
  {\cal A}(\mACSq,\mBCSq) = \sum_{j=1}^{N} c_j F_j(\mACSq,\mBCSq) \,,
\end{equation}
where $F_j(\mACSq,\mBCSq)$ are the dynamical amplitudes and $c_j$ are complex
coefficients describing the relative magnitude and phase of the different
decay channels. 
All the weak phase dependence is contained in the $c_j$ coefficients, which we
express in terms of their real and imaginary parts: $c_j = x_j + i y_j$, 
so $F_j(\mACSq,\mBCSq)$ contains kinematics and strong dynamics only.
We treat the $\Dzb\to\Kp\pim$ decay as flavour-specific and neglect
contributions from the doubly-Cabibbo-suppressed $\Dz\to\Kp\pim$ decay.
We assume direct \CP\ violation is negligible and hence use the same model for
$\Bz\to\Dzb\pipi$ and its conjugate decay. 
We also neglect possible contributions from $b \to u$ mediated, and hence
highly suppressed, transitions ({\it e.g.} $\Bz\to\DstarIIp\pim$).

In the $D\pi$ spectrum previous
studies~\cite{Kuzmin:2006mw,Aubert:2009wg,Abe:2003zm} have observed
contributions from \DstarII\ and \DstarI, as well as the effect of a virtual
\Dstar\ (\Dstarv) amplitude. 
The latter amplitude is described as virtual since although the region around
the narrow $\Dstar(2010)$ pole is vetoed, off-shell production can contribute
to the amplitude -- the effect is similar to a nonresonant P-wave term.
We find that an additional nonresonant (S-wave) $D\pi$ contribution is
necessary to fit the data;
we describe the nonresonant (NR) term using an empirical shape, first
introduced in Ref.~\cite{Garmash:2004wa}, proportional to $e^{-i\alpha \mBCSq}$,
where the shape parameter is determined from the data to be 
$\alpha = 0.60 \pm 0.15$ (statistical uncertainty only).
In the \pipi\ spectrum previous studies~\cite{Kuzmin:2006mw} have observed
contributions from \rhoIz\ and \fII. 
We find it is necessary to include S-wave terms and hence include a
contribution using the K-matrix 
formalism~\cite{Chung:1995dx,Aitchison:1972ay,Anisovich:2002ij},
described in more detail in the Appendix. 
To our knowledge, this is the first use of the K-matrix formalism in \B\ meson
decays. 
All other resonances are described using relativistic Breit--Wigner (RBW) shapes,
with Blatt--Weisskopf barrier form factors~\cite{blatt-weisskopf} and angular
distributions given in the Zemach tensor
formalism~\cite{Zemach:1963bc,Zemach:1968zz}.  The Dalitz plot formalism used
in this analysis is the same as that described in more detail in several
previous publications~\cite{Aubert:2005sk,Aubert:2005ce,Aubert:2008bj,:2009az}. 
The masses and widths of all resonances are constrained to world-average
values~\cite{Amsler:2008zz}, while K-matrix parameters are fixed to the values
tabulated in the Appendix.

In total there are 43 free parameters of the fit.
These are the yields of signal, \qqbar\ and the 6 \BB\ background categories;
the real and imaginary parts of 5 intermediate contributions to the signal DP
model (not counting those of $\DstarIIm\pip$ which are fixed as a reference);
the real and imaginary parts of 10 complex coefficients in the 
production vector of the K-matrix parametrization of the \pipi\ S-wave; 2
parameters each of the CR signal \DeltaE\ and \mes\ PDFs and the slope of the
continuum \DeltaE\ distribution.

\section{Results}

\begin{figure}[htb]
  \begin{center}
    \includegraphics[width=0.45\textwidth]{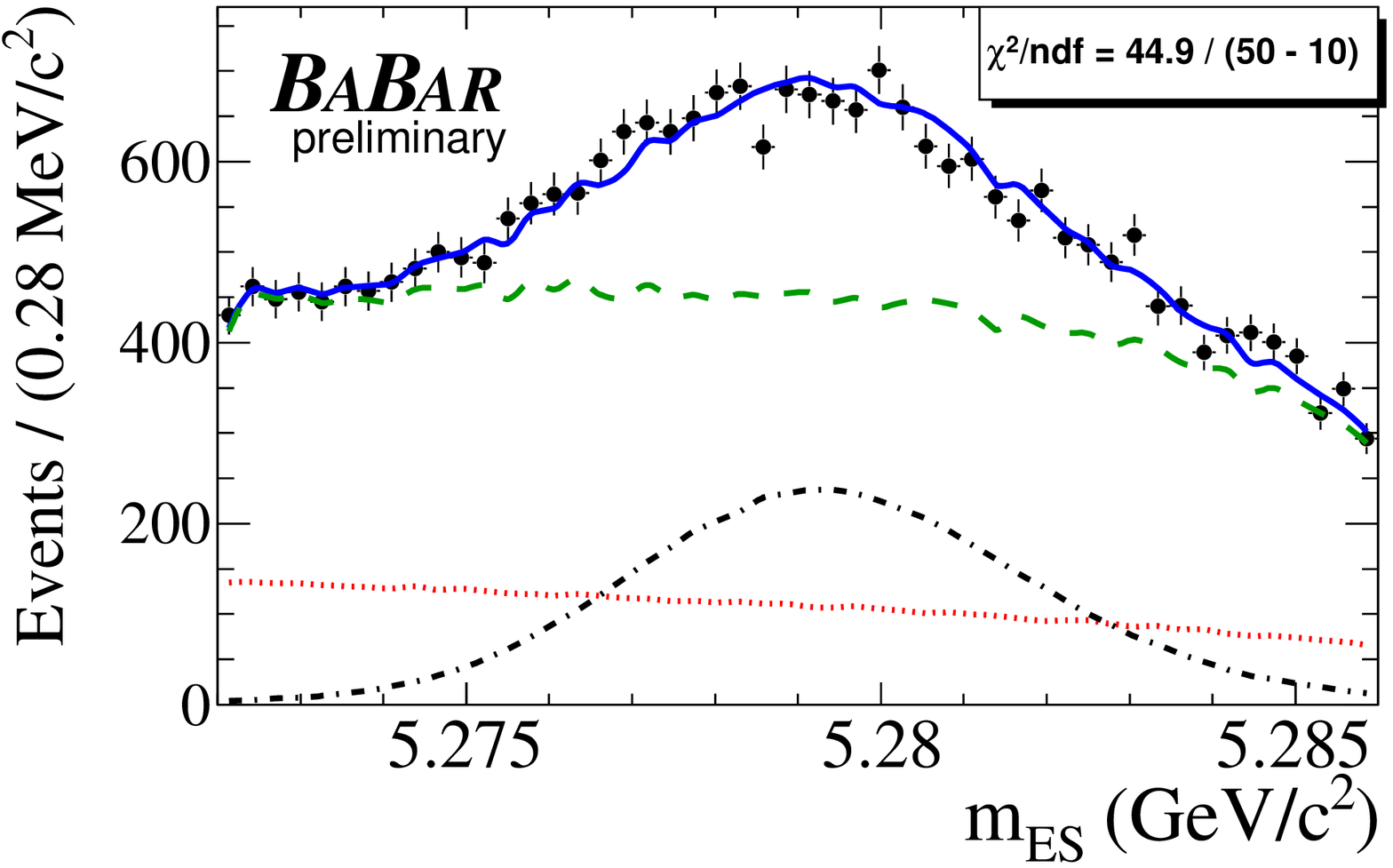}
    \includegraphics[width=0.45\textwidth]{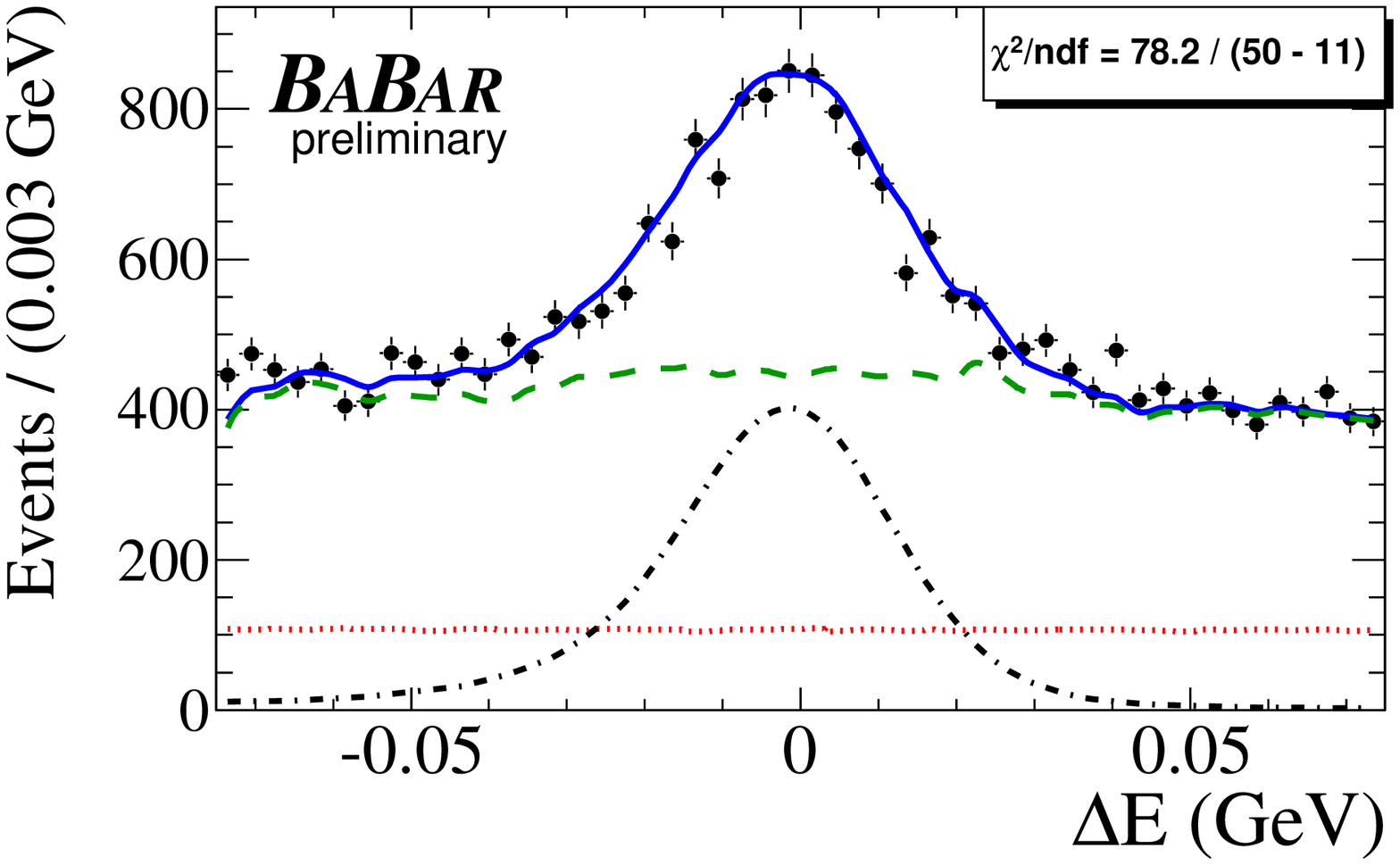}
    \caption{
      \mes\ and \DeltaE\ distributions. 
      The points with error bars show the data, the red dotted lines show the
      continuum background, the green dashed lines show the total background,
      the black dot-dashed lines show the signal, and the blue solid lines
      show the total fit result.}
    \label{fig:mes-de}
  \end{center}
\end{figure}

\begin{figure}[htb]
  \begin{center}
    \includegraphics[width=0.32\textwidth]{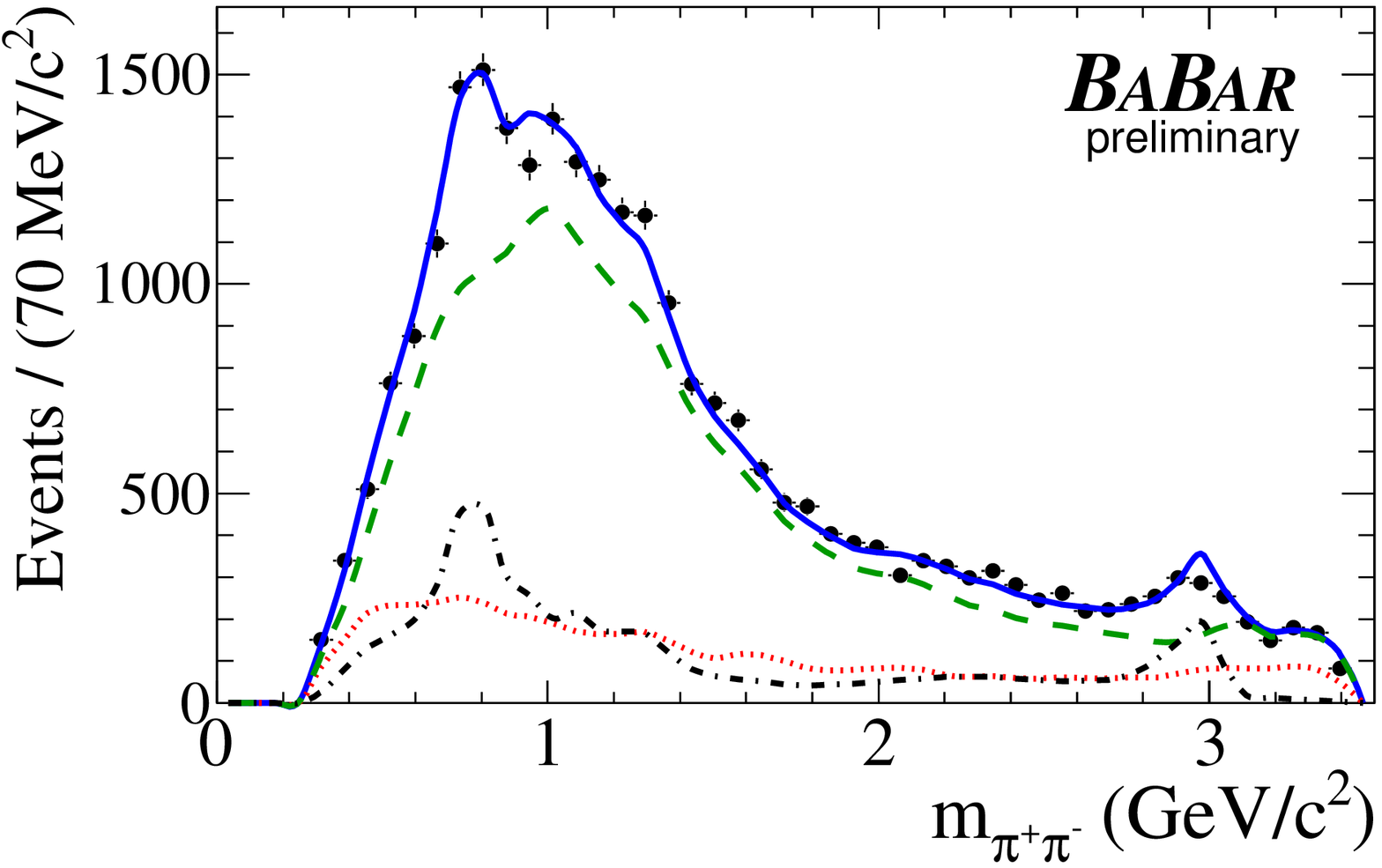}
    \includegraphics[width=0.32\textwidth]{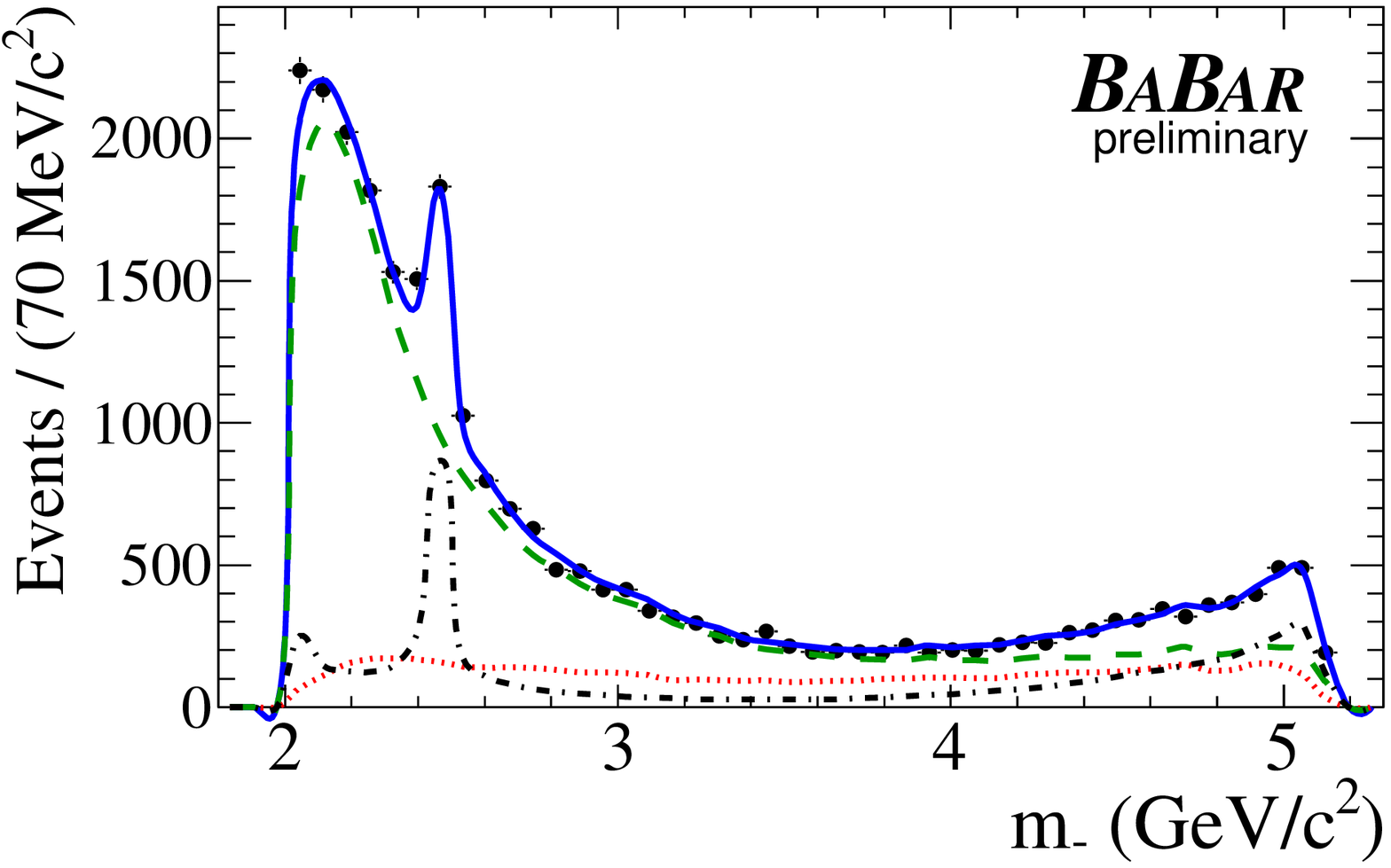}
    \includegraphics[width=0.32\textwidth]{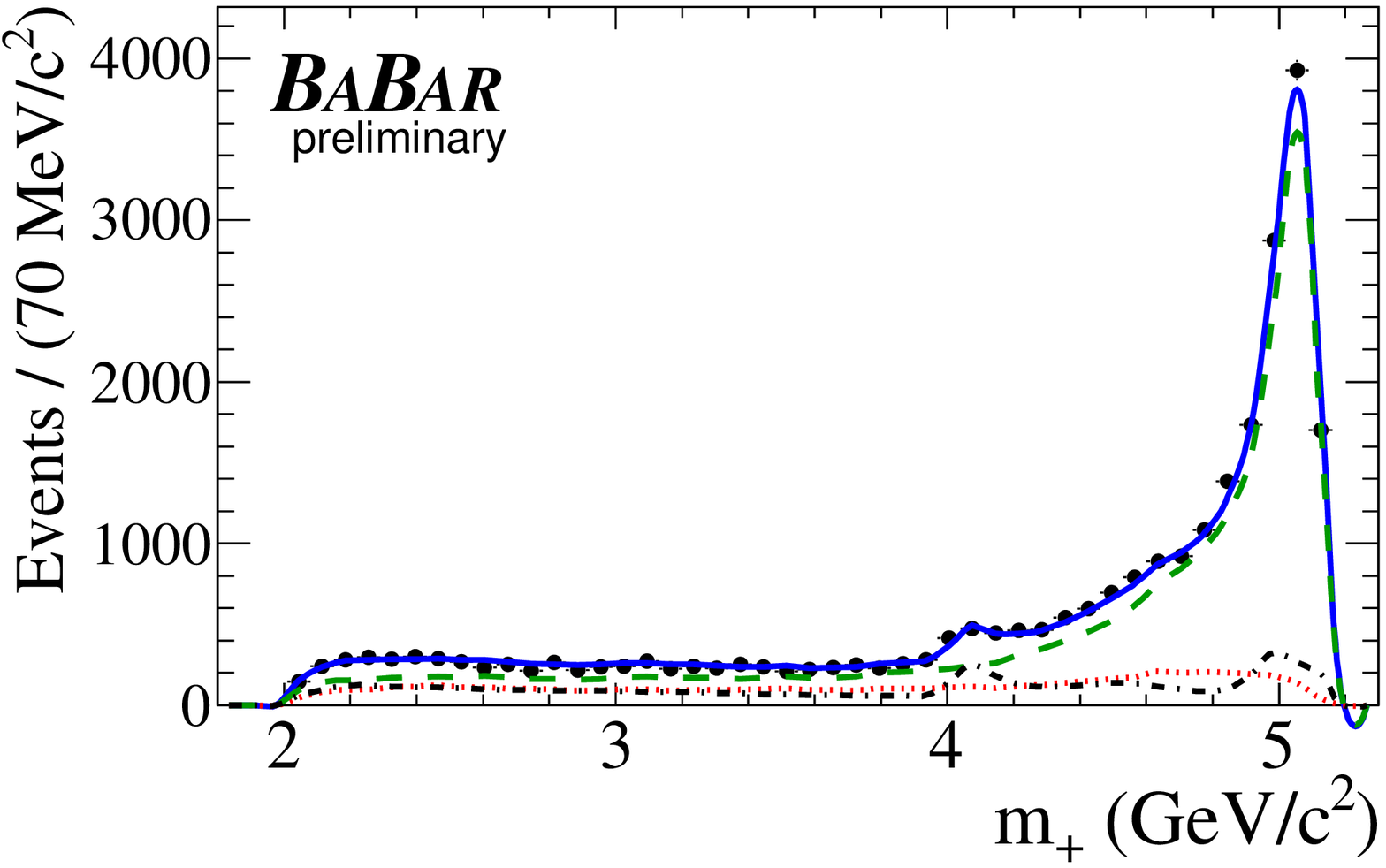} \\
    \includegraphics[width=0.32\textwidth]{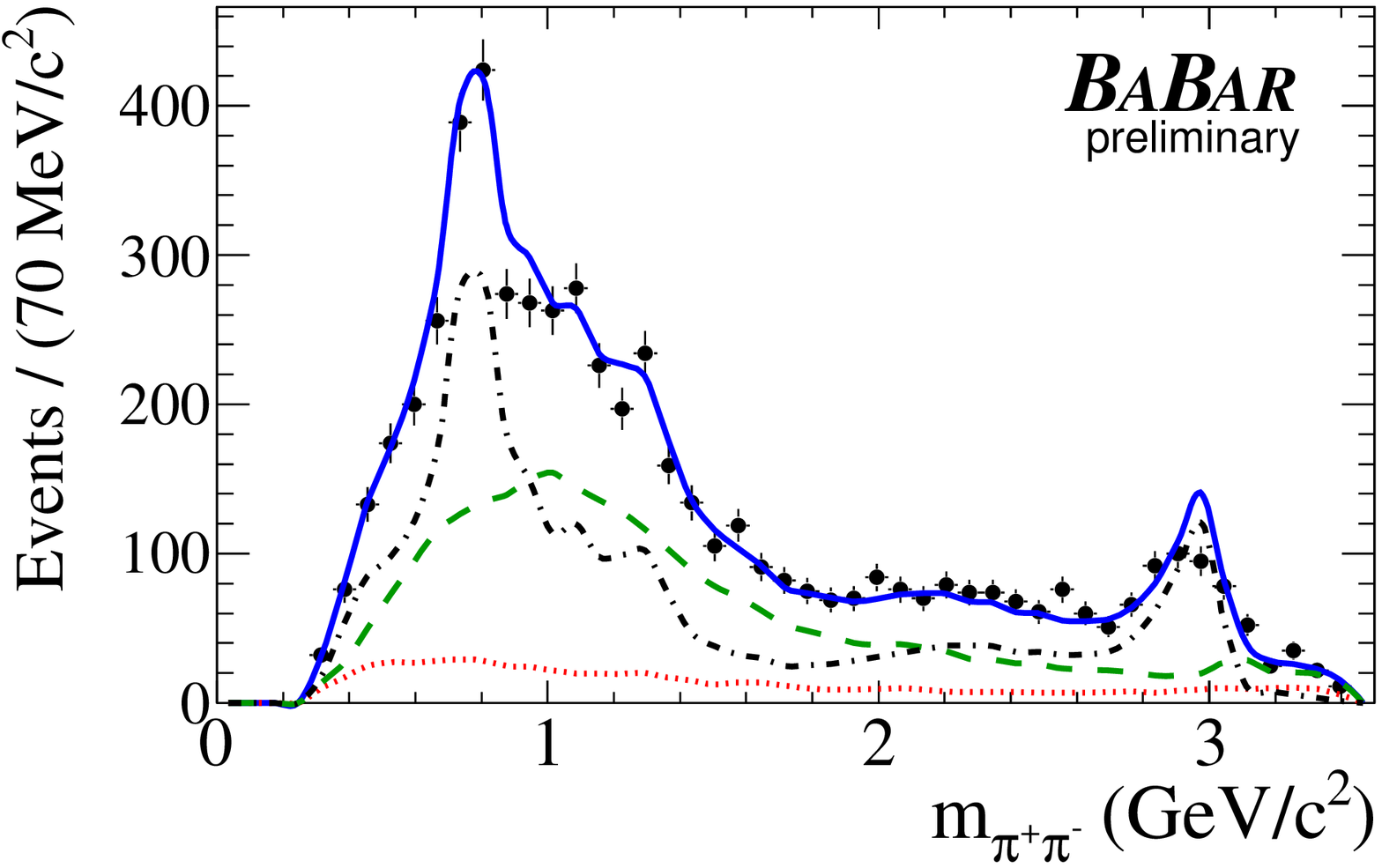}
    \includegraphics[width=0.32\textwidth]{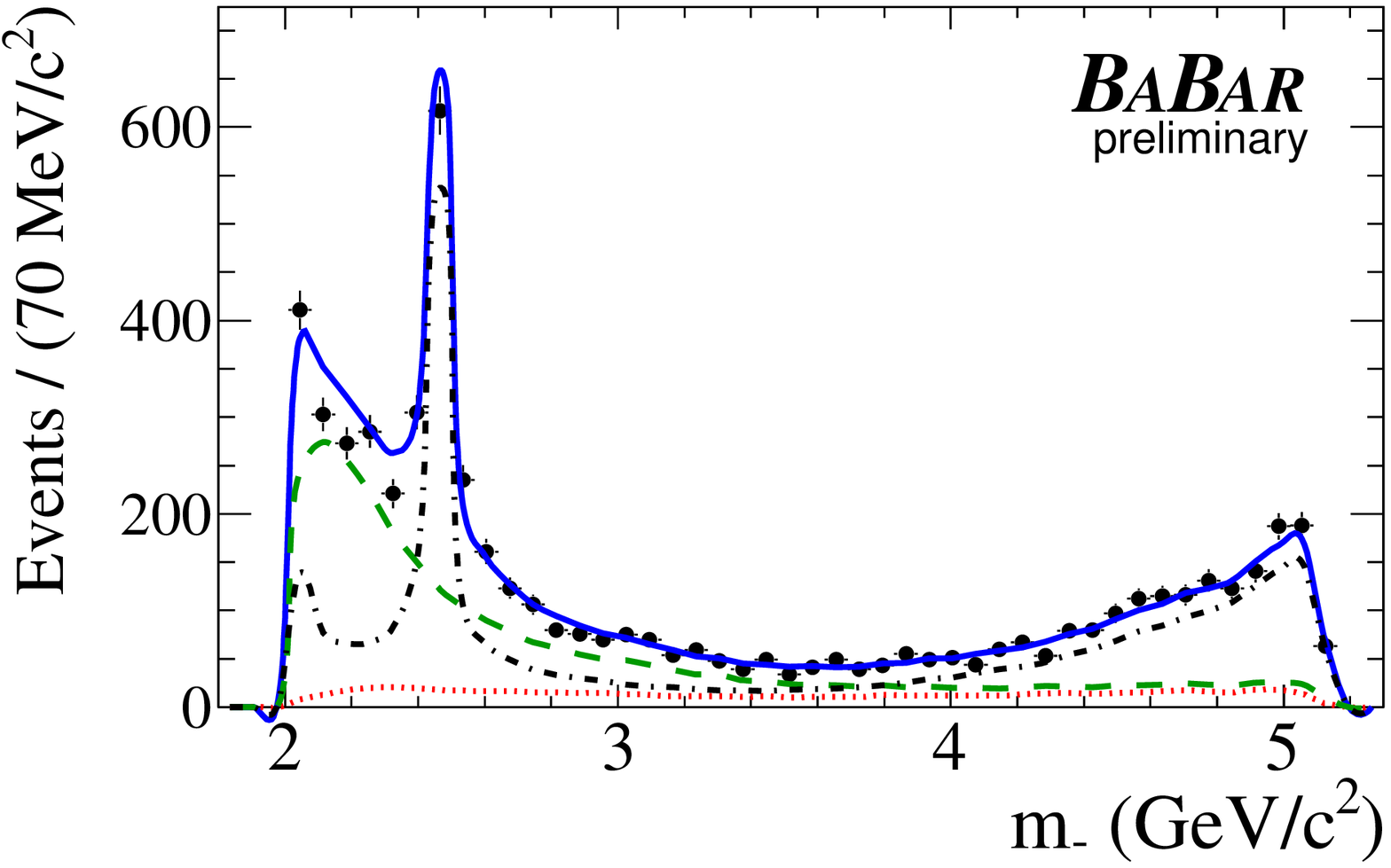}
    \includegraphics[width=0.32\textwidth]{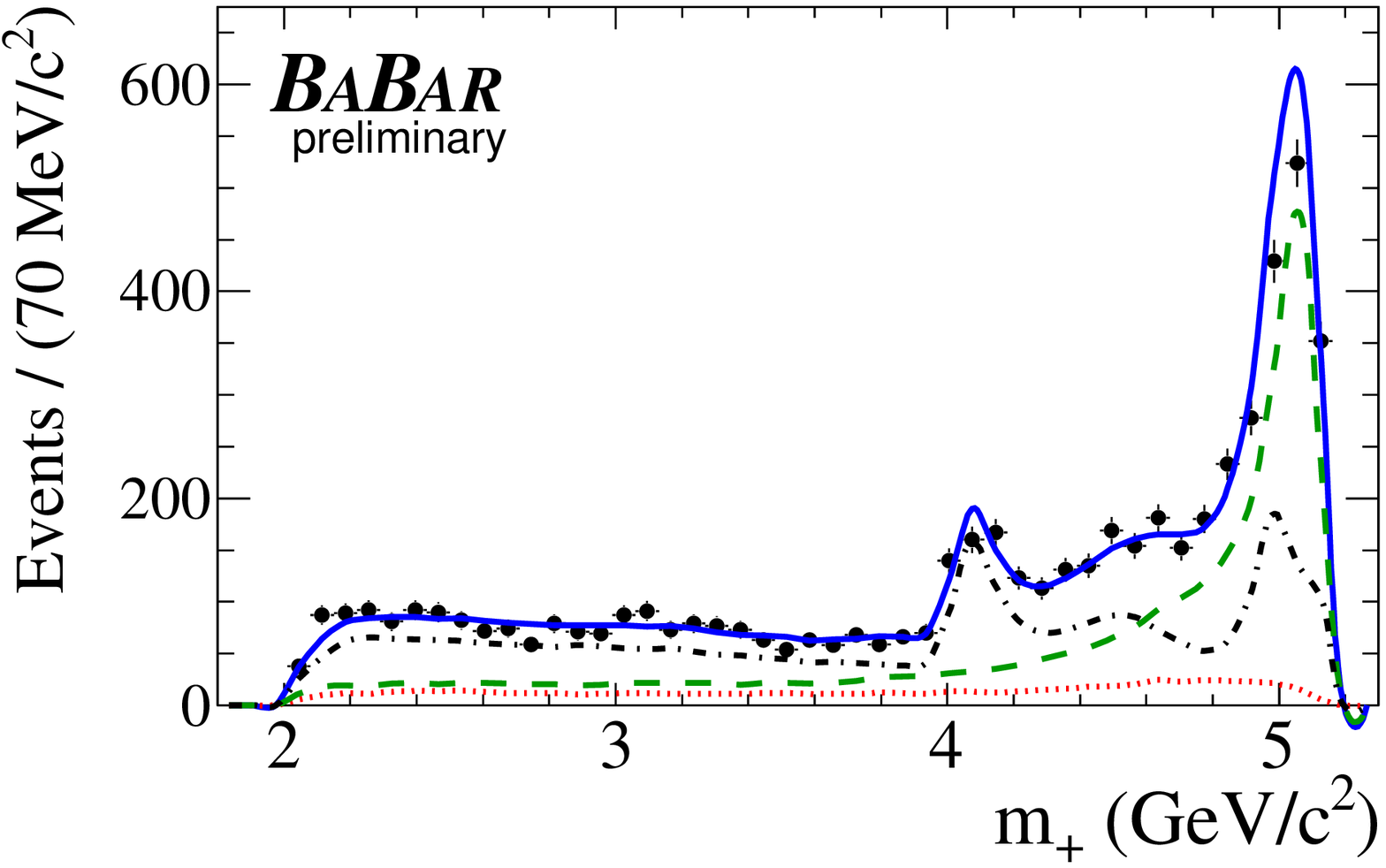}
    \caption{
      Projections of the fit results onto the invariant masses.
      The top row shows the projection of all the data; 
      the bottom row has the signal component enhanced by additional
      requirements  
      ($5.276\gevcc < \mes < 5.282\gevcc$ and $\abs{\DeltaE}<20\mev$).
      The points with error bars show the data, the red dotted lines show the
      continuum background, the green dashed lines show the total background,
      the black dot-dashed lines show the signal, and the blue solid lines
      show the total fit result.
    }
    \label{fig:dp-projections}
  \end{center}
\end{figure}

\begin{figure}[htb]
  \begin{center}
    \includegraphics[width=0.32\textwidth]{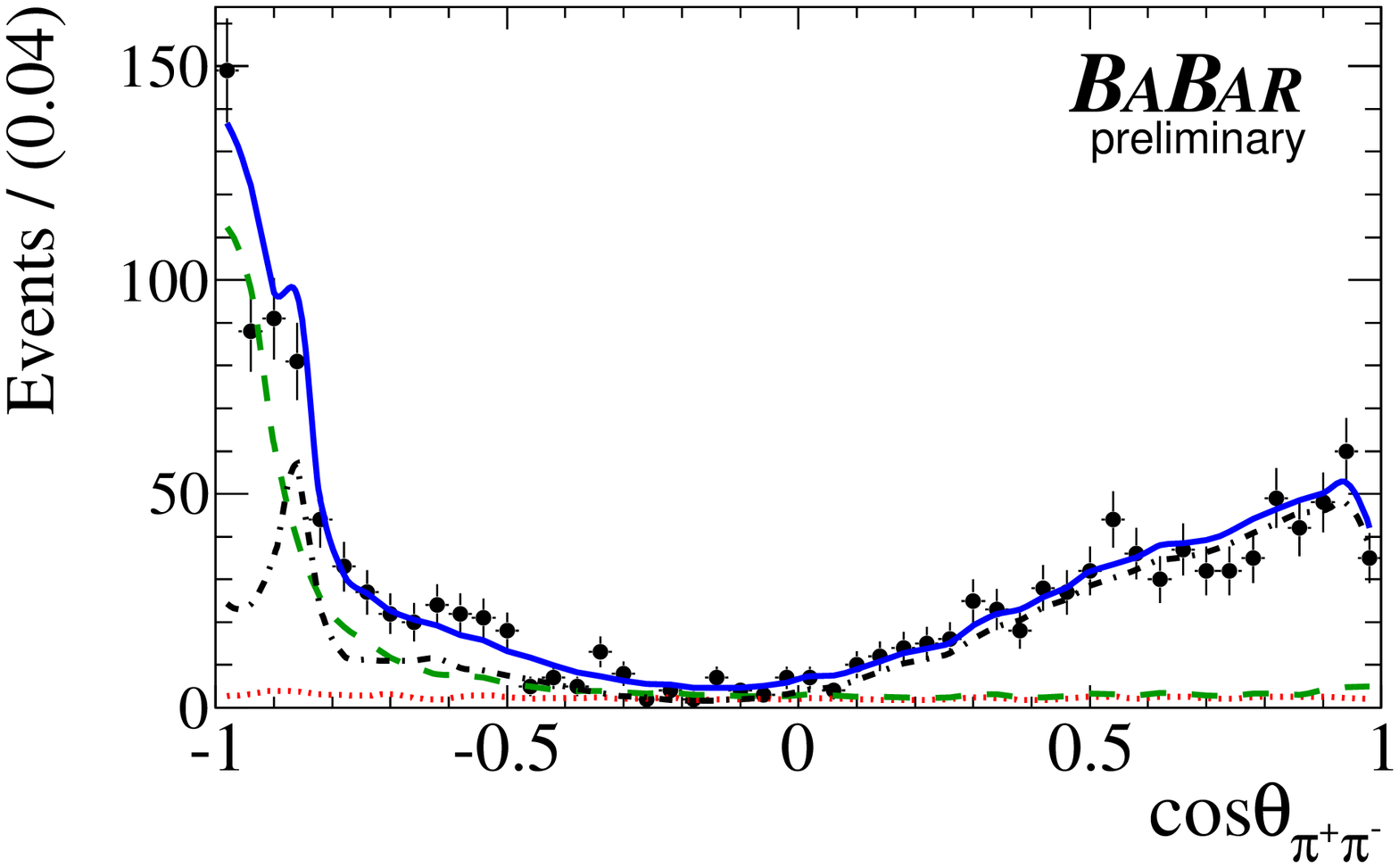}
    \includegraphics[width=0.32\textwidth]{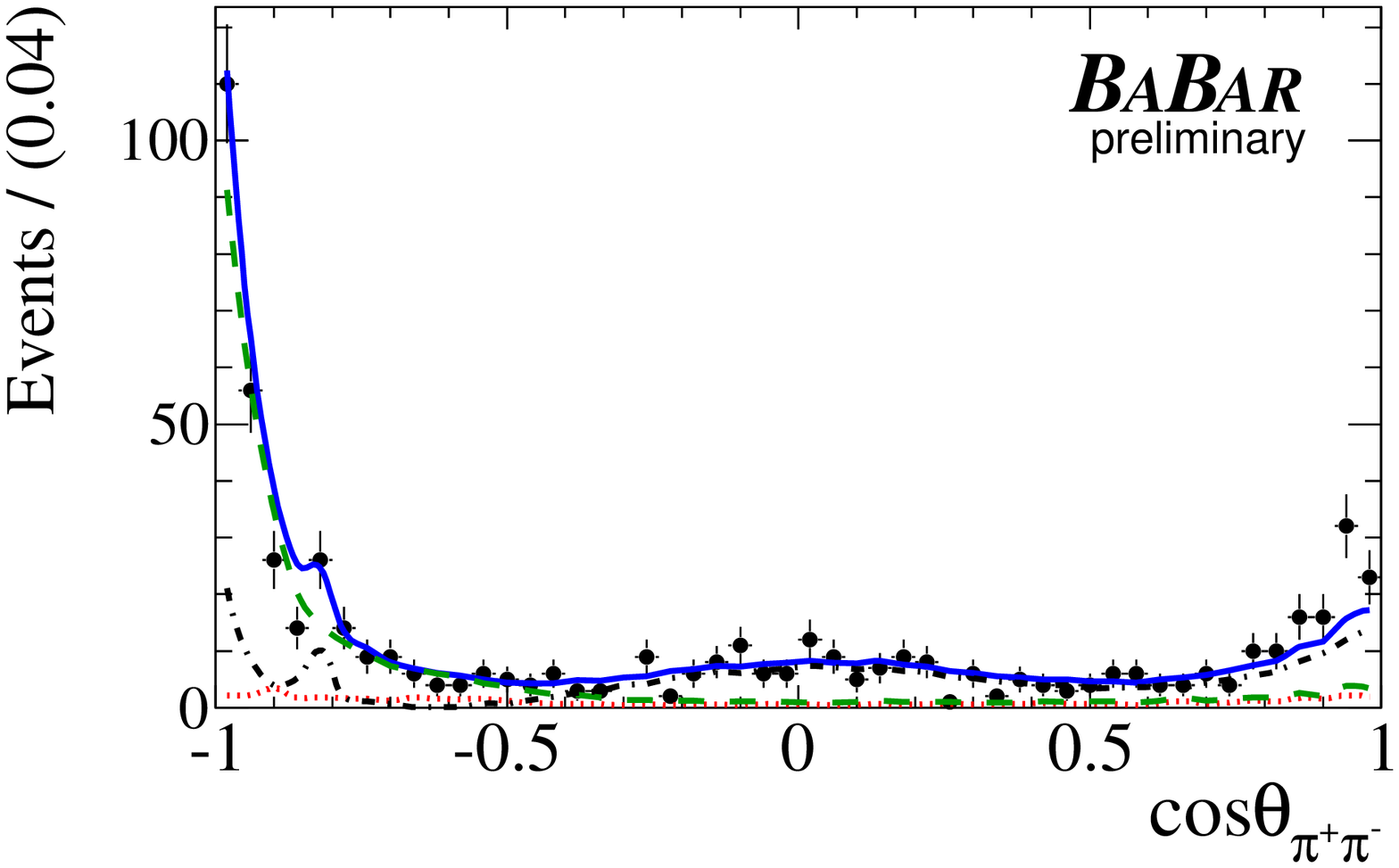}\\
    \includegraphics[width=0.32\textwidth]{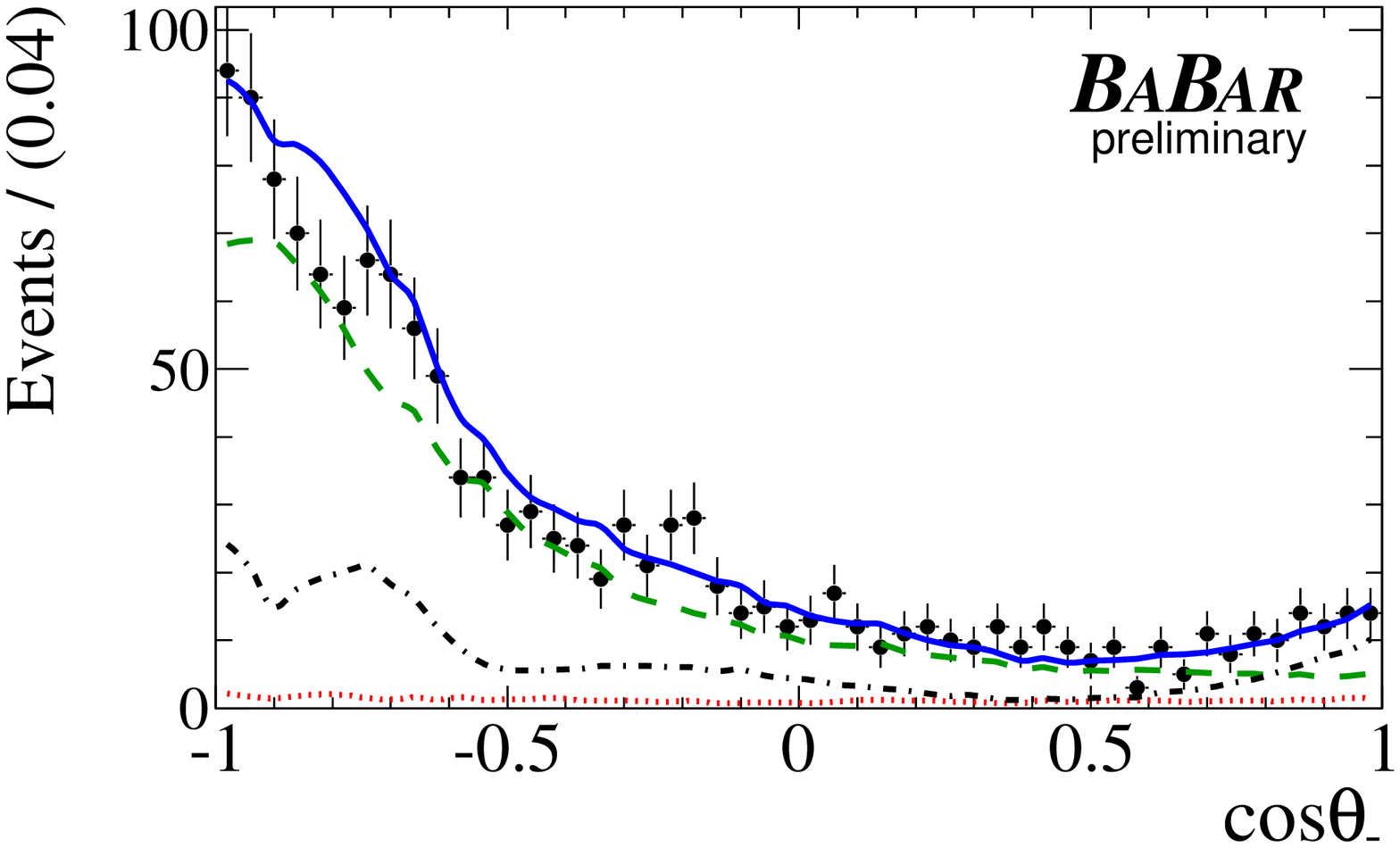}
    \includegraphics[width=0.32\textwidth]{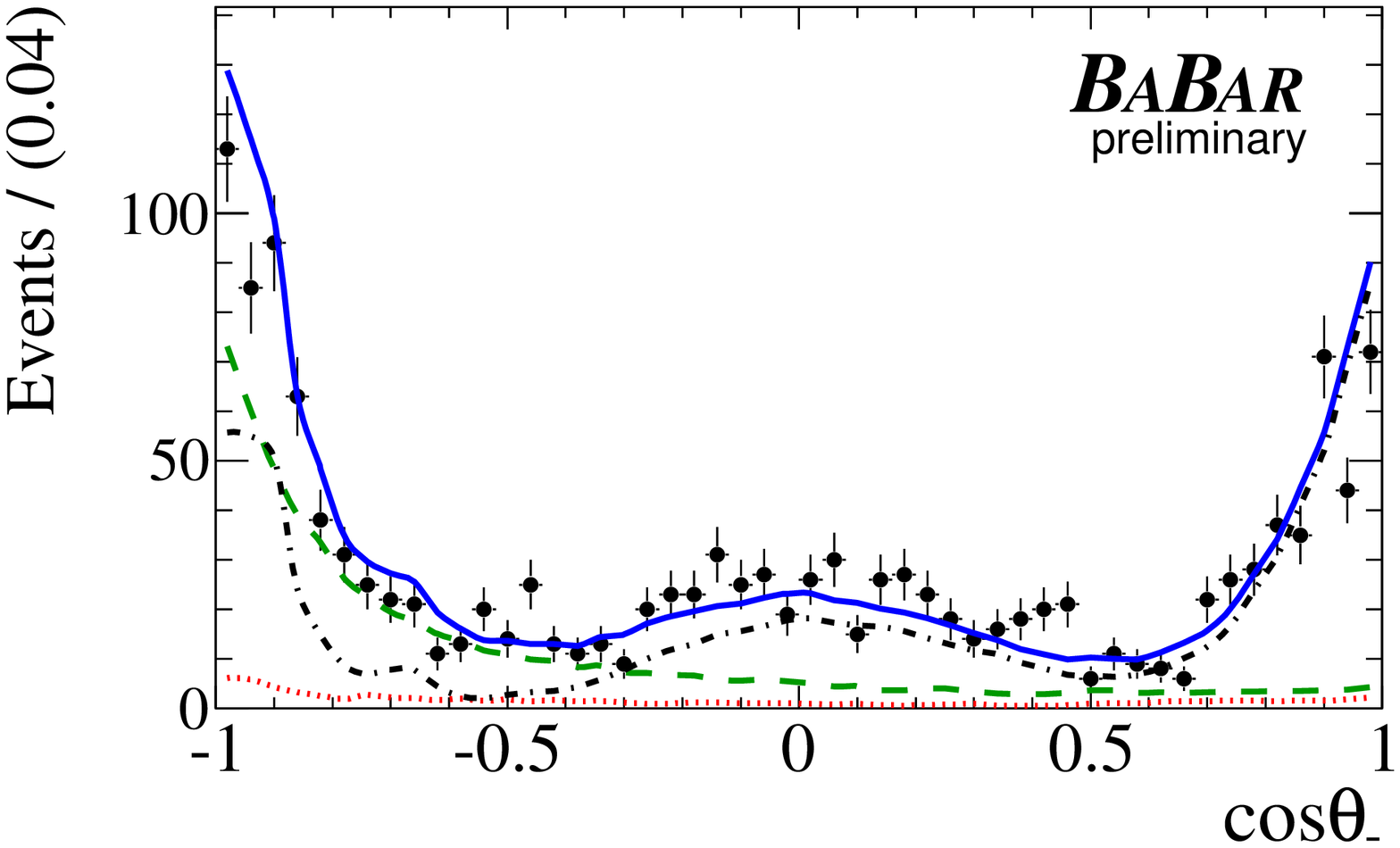}
    \caption{
      Projections onto cosines of helicity angles.
      The signal component has been enhanced in all plots by additional cuts
      ($5.276\gevcc < \mes < 5.282\gevcc$ and $\abs{\DeltaE}<20\mev$).
      The top row shows the projection onto the cosine of the \pipi\ helicity
      angle in the regions
      (left) around the \rhoIz, and (right) around the \fII.
      The bottom row shows the projection onto the cosine of the $D\pi$
      helicity angle in the regions (left) below
      and (right) around the \DstarII.
      The points with error bars show the data, the red dotted lines show the
      continuum background, the green dashed lines show the total background,
      the black dot-dashed lines show the signal, and the blue solid lines
      show the total fit result.}
    \label{fig:helicity-projections}
  \end{center}
\end{figure}

\begin{figure}[htb]
  \begin{center}
    \includegraphics[width=0.45\textwidth]{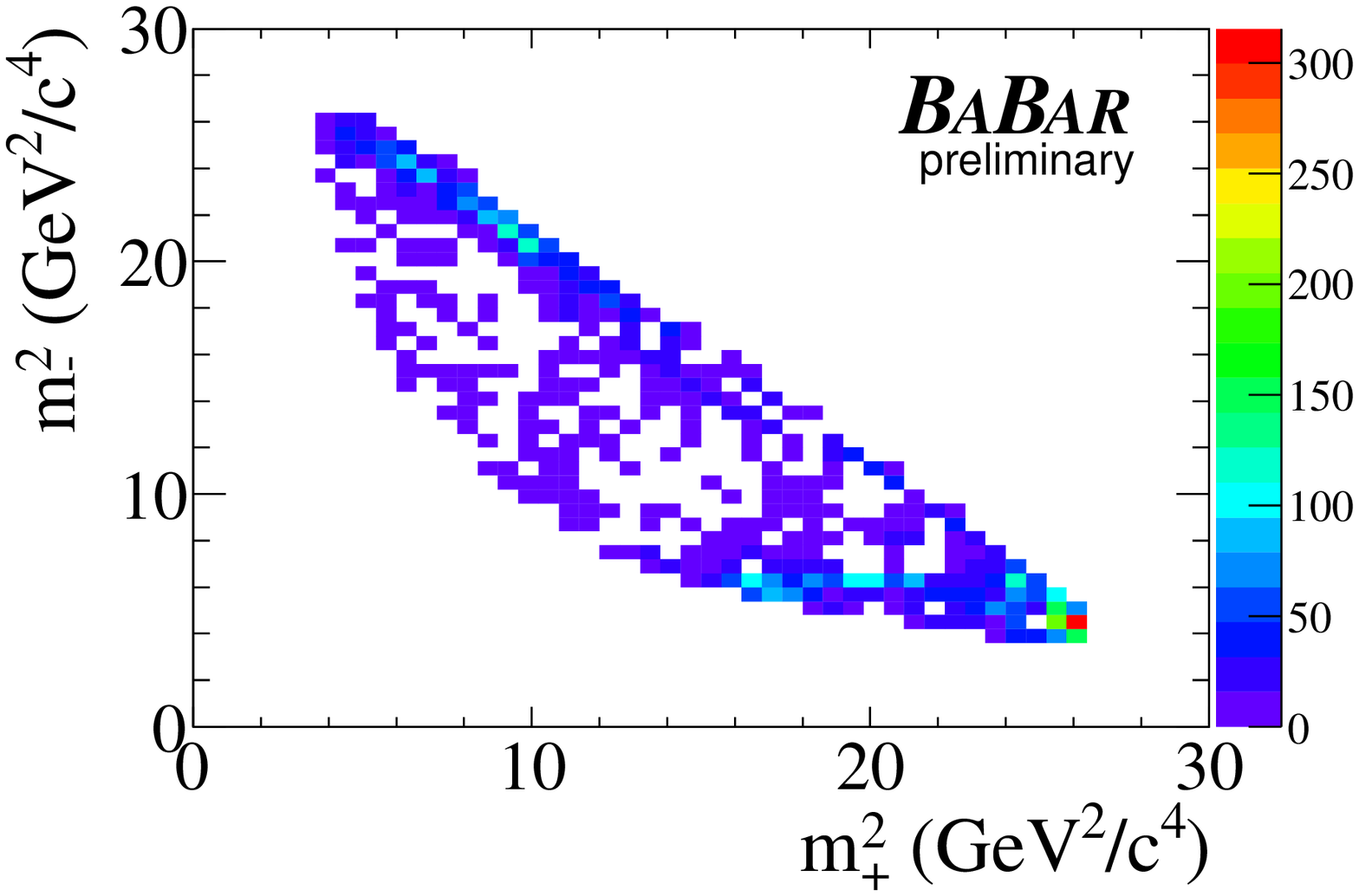}
    \includegraphics[width=0.45\textwidth]{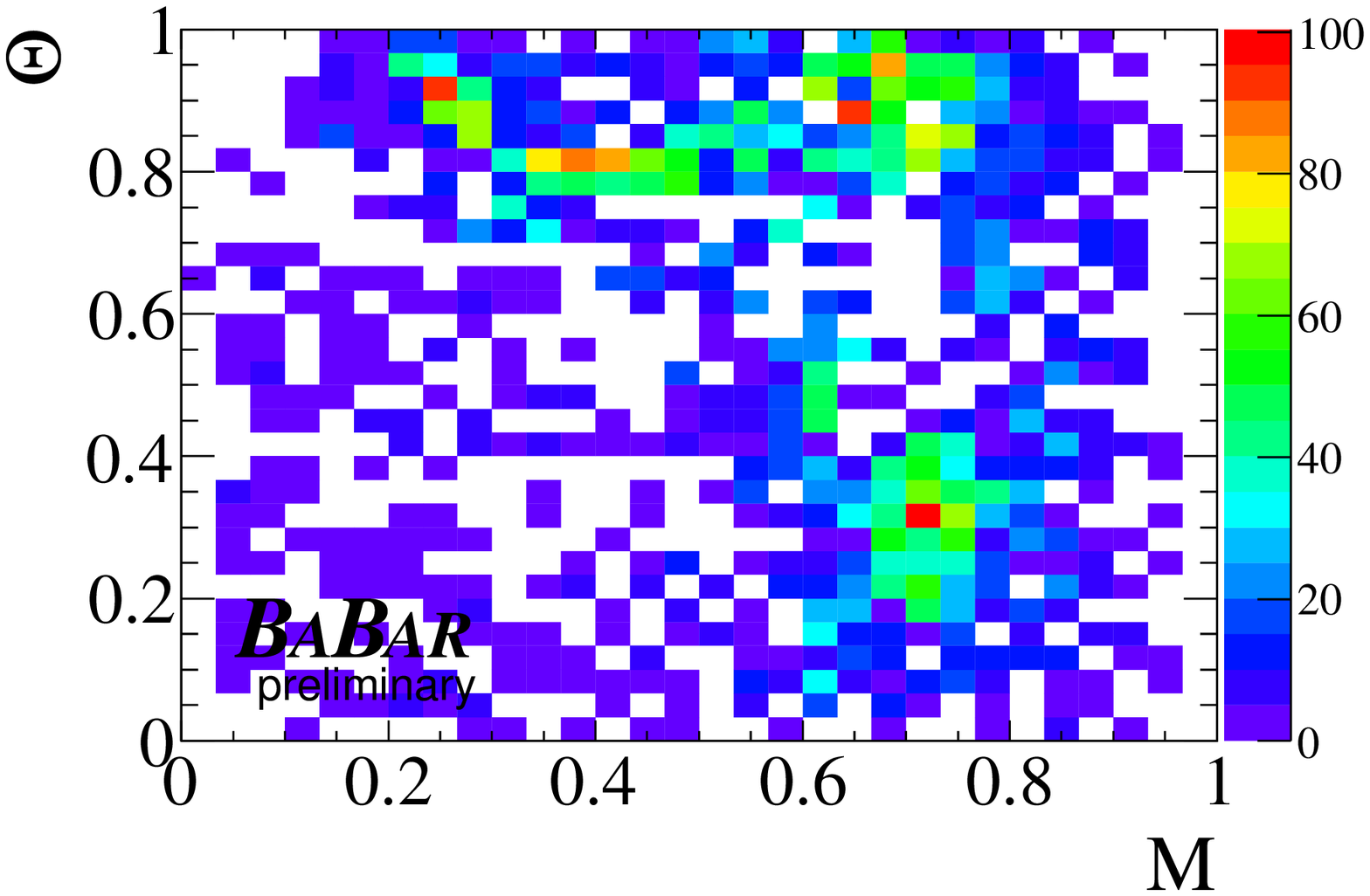}
    \caption{
      \splots\ of the signal distribution in the 
      (left) Dalitz plot and (right) square Dalitz plot.
    }
    \label{fig:dp-sdp-splots}
  \end{center}
\end{figure}

The fit returns \nsig\ signal events.
For this and all other quantities the statistical uncertainties are calculated
from an MC study where the events are generated from the PDFs and the PDF
parameters are the central values from the fit to data.
Yields of the various background categories are broadly in line with
expectation, although there appears to be some cross-feed between 
\BB\ categories.
Projections of the fit result onto \mes\ and \DeltaE\ are shown in
\figref{mes-de}, while projections onto each of the two-particle invariant
masses are shown in \figref{dp-projections} and projections onto the cosines
of the helicity angles, defined as the direction of one of the two daughters
of the resonance relative to the direction of the third particle in the rest
frame of the resonance, are shown in \figref{helicity-projections}.
The signal distribution across the phase-space, in both conventional and
square Dalitz plot co-ordinates, calculated using the 
\splot\ technique~\cite{Pivk:2004ty}, is shown in \figref{dp-sdp-splots}.
Structures due to the \DstarIIm, \rhoIz\ and \fII\ resonances are clearly
visible. 

Figures~\ref{fig:dp-projections}~and~\ref{fig:helicity-projections} show that
our DP model gives an excellent representation of the data in most regions of
the Dalitz plot.  The only region where discrepancies between the data and the
fit result are apparent is at low values of \mBC, where a sharp rise near
threshold is observed.  This structure also appears as a reflection in the
\mAC\ and \cBC\ distributions.  We discuss this further when we consider
model uncertainties, below.

We calculate the fit fractions and
interference fit fractions, shown as a matrix in \tabref{ff-matrix}.
The fit fractions are the elements along the diagonal, and are given by
\begin{equation}
  {\it FF}_j =
  \frac
  {\int_{\rm DP}\left|c_j F_j(\mACSq,\mBCSq)\right|^2 d(\mACSq)d(\mBCSq)}
  {\int_{\rm DP}\left|\sum_j c_j F_j(\mACSq,\mBCSq)\right|^2 d(\mACSq)d(\mBCSq)}\, ,
\label{eq:fitfraction}
\end{equation}
while the interference fraction are the off-diagonal elements and are given by
\begin{equation}
  {\it FF}_{ij} =
  \frac
  {\int_{\rm DP} 2 \, {\rm Re}\left[c_ic_j^* F_i(\mACSq,\mBCSq)F_j^*(\mACSq,\mBCSq)\right] d(\mACSq)d(\mBCSq)}
  {\int_{\rm DP}\left|\sum_j c_j F_j(\mACSq,\mBCSq)\right|^2 d(\mACSq)d(\mBCSq)}\, ,
  \label{eq:intfitfraction}
\end{equation}
for $i<j$ only.
Note that, with this definition, ${\it FF}_{jj} = 2 {\it FF}_{j}$.
These give a convention independent representation of the population of the DP.
Although the sum of fit fractions can be greater than unity -- 
in this case it is $(148\pm5)\,\%$ (statistical uncertainty only) -- 
the sum including interference fit fractions must be identically equal to one.
The largest interference effect is between \DstarImpip\ and the $D\pi$
nonresonant amplitude.

\begin{table}
  \caption{Matrix of fit fractions and interference fractions (central values only without uncertainties).}
  \label{tab:ff-matrix}
\begin{tabular}{l| c c c c c c c}
\hline
\hline
            & \DstarIIm\pip & \DstarIm\pip & \rhoIzDzb  & \fIIDzb   & \Dstarvm\pip & $D\pi$ NR & K matrix  \\
\hline                                                           
\DstarIIm\pip & $ 0.2047$   & $\cdots$   & $\cdots$  & $\cdots$  & $\cdots$   & $\cdots$  & $\cdots$  \\
\DstarIm\pip  & $ 0.0000$   & $ 0.2481$  & $\cdots$  & $\cdots$  & $\cdots$   & $\cdots$  & $\cdots$  \\
\rhoIzDzb    & $-0.0133$   & $ 0.0264$  & $ 0.3343$ & $\cdots$  & $\cdots$   & $\cdots$  & $\cdots$  \\
\fIIDzb     & $-0.0130$   & $ 0.0223$  & $ 0.0000$ & $ 0.0983$ & $\cdots$   & $\cdots$  & $\cdots$  \\
\Dstarvm\pip  & $ 0.0000$   & $-0.0001$  & $-0.0565$ & $-0.0347$ & $ 0.1579$  & $\cdots$  & $\cdots$  \\
$D\pi$ NR   & $ 0.0000$   & $-0.2471$  & $-0.0246$ & $-0.0458$ & $ 0.0001$  & $ 0.1844$ & $\cdots$  \\
K matrix    & $ 0.0019$   & $-0.0672$  & $ 0.0000$ & $-0.0003$ & $-0.0016$  & $-0.0303$ & $ 0.2559$ \\
\hline
\hline
\end{tabular}
\end{table}

In \tabref{bf-results} we give results for the branching fractions.
The inclusive \BztoDzbpipi\ branching fraction is calculated by dividing the
signal yield by the average efficiency determined from the nominal model, by
the number of \BB\ pairs in the data sample,
and by the branching fraction for the $D$ decay 
(${\cal B}(\Dzb\to\Kp\pim) = (3.91 \pm 0.05) \times 10^{-2}$~\cite{Amsler:2008zz}).
The average efficiency is found to be $30.6\,\%$ and is further corrected for
the measured data/MC differences (discussed under systematic uncertainties
below).
Our result compares well to that of Belle: ${\cal B}(\BztoDzbpipi) = (8.4 \pm
0.4 \pm 0.8) \times 10^{-4}$~\cite{Kuzmin:2006mw}.
The product branching fractions for the contributing decay modes are obtained
by multiplying the inclusive branching fraction by the relevant fit fraction.
Where possible, these have also been corrected for subdecay branching
fractions (${\cal B}(\rhoIz\to\pipi)=(98.9\pm0.16)\%$,
${\cal B}(\fII\to\pipi)=(84.8_{-1.2}^{+2.4})\%$~\cite{Amsler:2008zz}).
We are not able to perform such a calculation for $\DstarIImpip$ since,
although decay modes other than $D\pi$ have been seen, the relative branching
fractions are not known.
The \DstarI\ has only been observed to decay into $D\pi$, but it may be
presumptuous to conclude that its branching fraction is 100\,\%.
Our results for \DstarIIm\pip, \rhoIzDzb\ and \fIIDzb\ are consistent with
those of Belle, while we see a somewhat larger branching fraction for
\Dstarvm\pip\ and a much larger branching fraction for \DstarIm\pip\ (Belle
measures ${\cal B}(\Bz\to\DstarIm\pip)\times{\cal B}(\DstarIm\to\Dzb\pim) = (0.60 \pm 0.13 \pm 0.15 \pm 0.22)\times 10^{-4}$).

\begin{table}
  \caption{Branching fraction results from the fit to data. The first
  uncertainty is statistical, the second is systematic, the third is due to
  the Dalitz-plot model, and the fourth (where present) is due to secondary
  branching fractions.
  The third column gives the product of the branching fraction of the $B$
  decay to the mode listed in the leftmost column with that of the
  intermediate resonance decay to the final state particles.}
  \label{tab:bf-results}
\begin{tabular}{l@{\hspace{5mm}} r@{$\pm$}c@{$\pm$}c@{$\pm$}l @{\hspace{5mm}} r@{$\pm$}c@{$\pm$}c@{$\pm$}c@{$\pm$}l @{\hspace{5mm}} r@{$\pm$}c@{$\pm$}c@{$\pm$}c@{$\pm$}l}
\hline
\hline
Resonance              & \fcc{Fit Fraction}                  & \fvcc{$\BR(\Bz\to{\rm Mode})$}             & \fvcc{$\BR(\Bz\to{\rm Mode})$}             \\
                       & \fcc{(\%)}                          & \fvcc{$\times\BR(R \to hh)$ ($10^{-4}$)}   & \fvcc{($10^{-4}$)}                         \\
\hline                                                           
Inclusive \BztoDzbpipi & \fcc{$\cdots$}                      & \fvcc{$\cdots$}                            & $8.81$ & $0.18$ & $0.76$ & $0.78$ & $0.11$ \\
\hline                                                           
\DstarIImpip           & $20.5$ & $0.9$ & $1.3$ & $ 3.7$ & $1.80$ & $0.09$ & $0.19$ & $0.37$ & $0.02$ & \fvcc{$\cdots$}                            \\
\DstarImpip            & $24.8 $ & $2.5 $ & $3.0 $ & $12.9 $ & $2.18$ & $0.23$ & $0.33$ & $1.15$ & $0.03$ & \fvcc{$\cdots$}                            \\
\rhoIzDzb               & $33.4 $ & $2.0 $ & $5.2 $ & $10.0 $ & $2.94$ & $0.19$ & $0.53$ & $0.92$ & $0.04$ & $2.98$ & $0.19$ & $0.53$ & $0.93$ & $0.04$ \\
\fIIDzb                & $ 9.8 $ & $1.1 $ & $1.6 $ & $ 3.4 $ & $0.86$ & $0.10$ & $0.16$ & $0.31$ & $0.01$ & $1.02$ & $0.12$ & $0.18$ & $0.36$ & $0.03$ \\
\Dstarvmpip            & $15.8$ & $0.9$ & $1.2$ & $ 3.7$ & $1.39$ & $0.08$ & $0.16$ & $0.35$ & $0.02$ & \fvcc{$\cdots$}                            \\
$D\pi$ nonresonant     & $18.4 $ & $2.3 $ & $4.3 $ & $13.6 $ & $1.62$ & $0.21$ & $0.41$ & $1.21$ & $0.02$ & \fvcc{$\cdots$}                            \\
K matrix total         & $25.6 $ & $2.5 $ & $3.2 $ & $ 6.1 $ & $2.26$ & $0.22$ & $0.34$ & $0.58$ & $0.03$ & \fvcc{$\cdots$}                            \\
\hline
\hline
\end{tabular}
\end{table}

\section{Systematic Uncertainties}

We consider the following systematic effects on the values of the fit fractions.
\begin{itemize}
\item Fixed shapes of the efficiency, \qqbar\ and \BB\ Dalitz-plot histograms: \\
  The contents of all bins of square Dalitz plot histograms used to describe these shapes are fluctuated in accordance with the uncertainties.  This procedure is repeated many times and the RMS of the distribution of the change in the fit results is taken as the associated systematic uncertainty.
\item Fixed \mes\ and \DeltaE\ PDF parameters (or histograms): \\
  We vary any fixed parameters in the PDF descriptions by their uncertainties, taking correlations into account.  The variation in the fit results is taken as the systematic uncertainty.
For most parameters, their values and uncertainties are determined from data
control samples.  An exception is the self-cross-feed fraction which is
obtained from Monte Carlo.  To conservatively allow for possible data/MC
differences in the behaviour of the SCF component, we apply a Dalitz-plot
independent scale factor that alternately increases and decreases the SCF
fraction by a factor of two, and take the larger difference compared to the
nominal result as the uncertainty.  The contents of the histograms used to
describe the \BB\ background \mes\ and \DeltaE\ PDFs are varied using the same
prescription as described above.
\item Fit bias: \\
  We generate large ensembles of pseudo-experiments, containing fully
  simulated signal events, using the parameters returned by the fit to
  data.  From the distribution of results of these ensembles, we evaluate
  biases on the fit parameters.  All biases are found to be small compared
  to the statistical uncertainties.  
We assign systematic uncertainties of the sum in quadrature of half the
bias and its uncertainty.
\end{itemize}
These sources of systematic uncertainty are summarized in
\tabref{systematics}.
The total is obtained by combining all sources in quadrature.

\begin{table}
  \caption[Yield and fit fraction systematics]
  {Systematic uncertainties on the signal yield and fit fractions.}
  \label{tab:systematics}
\resizebox{0.99\textwidth}{!}{
  \begin{tabular}{lcccccccccc}
    \hline
    \hline
    & Efficiency & \qqbar & \BB     & CR \mes\ \& \DeltaE 
    & SCF \mes\ \& \DeltaE & \BB\ \mes\ \& & SCF      
    & \qqbar\ \mes   & Fit bias & Total \\
    &            & DP PDF & DP PDFs & PDF parameters      
    & PDF parameters       & \DeltaE\ PDFs & fraction 
    & PDF parameters & \\
    \hline
    Signal yield	& 6.3	 & 19		 & 24		 & 35		 & 0.7		 & 22		 & 302		 & 2.1		 & 46		 & 310		\\
    \DstarIImpip\ FF	& 0.0018  & 0.0036	 & 0.0051	 & 0.00072	 & 0.00005	 & 0.0033	 & 0.010	 & 0.00009	 & 0.0037	 & 0.013	\\
    \DstarImpip\ FF	& 0.003	  & 0.016	 & 0.024	 & 0.00063	 & 0.00013	 & 0.0065	 & 0.00096	 & 0.00011	 & 0.0023	 & 0.030	\\
    \rhoIzDzb\ FF	& 0.016	  & 0.028	 & 0.031	 & 0.00069	 & 0.00010	 & 0.025	 & 0.0078	 & 0.00023	 & 0.0014	 & 0.052	\\
    \fIIDzb\ FF		& 0.0054  & 0.0091	 & 0.0077	 & 0.00040	 & 0.00005	 & 0.0078	 & 0.0029	 & 0.00006	 & 0.0011	 & 0.016	\\
    \Dstarvmpip\ FF	& 0.00097 & 0.0028	 & 0.0045	 & 0.00034	 & 0.00004	 & 0.0027	 & 0.0091	 & 0.00007	 & 0.0052	 & 0.012	\\
    $D\pi$ NR FF	& 0.015	  & 0.023	 & 0.022	 & 0.00052	 & 0.00010	 & 0.020	 & 0.015	 & 0.00008	 & 0.0036	 & 0.043	\\
    K matrix total FF	& 0.0057  & 0.014	 & 0.015	 & 0.00075	 & 0.00017	 & 0.012	 & 0.018	 & 0.00008	 & 0.010	 & 0.032	\\
    \hline
    \hline
  \end{tabular}
}
\end{table}

We consider additional systematic effects on the values of the branching fractions.
These are uncertainties on the differences between the efficiencies of
selection requirements on data and MC for tracking (1.0\%), particle
identification (4.0\%), the neural network cut (3.2\%), and the number of \BB\
pairs (0.6\%). 
Furthermore, where we have divided by a daughter branching fraction in order to isolate the \B\ decay branching fraction, any uncertainty in the world average value used in the division also contributes systematic uncertainty.

An additional source of uncertainty in Dalitz-plot analyses arises due to the composition of the Dalitz plot.
We consider the following sources of model uncertainty:
\begin{itemize}
\item Fixed parameters of contributing amplitudes: \\
  We vary the masses and widths of all resonances described by RBW shapes
  according to the uncertainties of the world average
  values~\cite{Amsler:2008zz} (with the exception of the $\Dstar_{0}(2400)$
  mass, which we vary by $\pm 100\mevcc$ to account for the discrepancy in the
  measured masses of charged and neutral isospin partners).  We vary the
  $\alpha$ parameter of the $D\pi$ nonresonant contribution within its
  uncertainty.  We change the radius parameter of the Blatt--Weisskopf factors
  from its nominal value of $4\gev^{-1}$ to both $3\gev^{-1}$ and $5\gev^{-1}$. 
\item Alternative parameterisations: \\
  We use the Gounaris--Sakurai lineshape~\cite{Gounaris:1968mw} as an
  alternative description for the \rhoI\ resonance.  We replace the \pipi\
  S-wave K-matrix term with contributions used in the analysis of
  $\Bz\to\Dzb\pipi$ by Belle~\cite{Kuzmin:2006mw}, namely $\sigma$ (described
  as in Ref.~\cite{Bugg:2003kj}), \fI\ (described by the Flatt\'e
  distribution~\cite{Flatte:1976xu}) and \fIII\ (RBW). 
  To address the possible discrepancy between the data and the fit result at
  low values of \mBC, we replace the $D\pi$ nonresonant contribution with a
  functional form proposed for a putative ``dabba'' state~\cite{Bugg:2009tu}.
  We have also performed a fit in which the background from
  $\Dstarm(2010)\pip$ events escaping the veto is treated as a separate
  (seventh) \BB\ background category, and include the deviation in the results
  as a source of model uncertainty.
\item Additional possible contributions: \\
  We repeat the fit adding states to the model: \omegaI, \rhoII, $D(2600)$ (both as scalar and vector) and $D(2760)$ (vector)~\cite{Aubert:201zzz}.
\end{itemize}
A summary is given in \tabref{model-errors}.
The total model uncertainty is obtained by combining all sources in quadrature.

\begin{table}
  \caption[Yield and fit fraction model uncertainties]
  {Dalitz-plot model uncertainties on the signal yield and fit fractions.
  Refer to the text for details of the model variations.}
  \label{tab:model-errors}
  \begin{tabular}{lccccccc}
    \hline
    \hline
			& Mass \& & $D\pi$ NR	 & BW barrier	 & \rhoIz\ GS	 & $D\pi$ S-wave & \pipi\ S-wave \\
			& width	  & $\alpha$	 & radius	 & lineshape	 & ``dabba''	 & 		 \\
    \hline                                                                                                                                                                       
    Signal yield	& 44	  & 6.4		 & 11		 & 1.1		 & 14		 & 67		 \\
    \DstarIImpip\ FF	& 0.028	  & 0.0027	 & 0.020	 & 0.00007	 & 0.0019	 & 0.0052	 \\
    \DstarImpip\ FF	& 0.061	  & 0.031	 & 0.0098	 & 0.00066	 & 0.099	 & 0.043	 \\
    \rhoIzDzb\ FF	& 0.045	  & 0.0056	 & 0.042	 & 0.0010	 & 0.00012	 & 0.034	 \\
    \fIIDzb\ FF		& 0.018	  & 0.00061	 & 0.0060	 & 0.00058	 & 0.0040	 & 0.014	 \\
    \Dstarvmpip\ FF	& 0.018	  & 0.0028	 & 0.015	 & 0.00076	 & 0.025	 & 0.0097	 \\
    $D\pi$ NR FF	& 0.10	  & 0.024	 & 0.021	 & 0.0060	 & $\cdots$	 & 0.026	 \\
    K matrix total FF	& 0.023	  & 0.0075	 & 0.010	 & 0.0034	 & 0.038	 & $\cdots$	 \\
    \hline
			& Add \omegaI	 & Add \rhoII	 & Add $D(2600)$ & Add $D(2600)$ & Add $D(2760)$ & 7 \BB\ Cat.	& Total	\\
			& 		 & 		 & (scalar)	 & (vector)	 & (vector)	 &  		& 	\\
    \hline                                                                                                 
    Signal yield	& 11		 & 53		 & 0.62		 & 13		 & 1.5		 & 8.4		& 100	\\
    \DstarIImpip\ FF	& 0.00019	 & 0.0060	 & 0.00026	 & 0.011	 & 0.00063	 & 0.00088	& 0.037	\\
    \DstarImpip\ FF	& 0.0019	 & 0.00091	 & 0.0085	 & 0.011	 & 0.0090	 & 0.00034	& 0.129	\\
    \rhoIzDzb\ FF	& 0.015		 & 0.047	 & 0.00075	 & 0.050	 & 0.0096	 & 0.0015	& 0.100	\\
    \fIIDzb\ FF		& 0.00048	 & 0.010	 & 0.00004	 & 0.021	 & 0.0021	 & 0.00050	& 0.034	\\
    \Dstarvmpip\ FF	& 0.0015	 & 0.0095	 & 0.00012	 & 0.0040	 & 0.0016	 & 0.00088	& 0.037	\\
    $D\pi$ NR FF	& 0.0053	 & 0.045	 & 0.0025	 & 0.065	 & 0.0036	 & 0.0012	& 0.136	\\
    K matrix total FF	& 0.0070	 & 0.034	 & 0.0019	 & 0.018	 & 0.011	 & 0.00073	& 0.061	\\
    \hline
    \hline
  \end{tabular}
\end{table}

\section{Discussion}

Isospin symmetry can be used to relate the decay amplitudes of
$\Bz\to\Dzb\rhoz$, $\Bz\to\Dm\rhop$ and
$\Bp\to\Dzb\rhop$~\cite{Rosner:1999zm,Neubert:2001sj,Chiang:2002tv,Mantry:2004pg}:
\begin{eqnarray}
  A(\Dzb\rhop) & = & \sqrt{3}A_{3/2} \, , \\
  A(\Dm \rhop) & = & \sqrt{1/3}A_{3/2} + \sqrt{2/3}A_{1/2} \, , \\
  \sqrt{2}A(\Dzb\rhoz) & = & \sqrt{4/3}A_{3/2} - \sqrt{2/3}A_{1/2} \, ,
\end{eqnarray}
where $A_{3/2}$ and $A_{1/2}$ are the amplitudes for isospin $3/2$ and $1/2$
final states respectively.
These equations give the triangle relation
\begin{equation}
  A(\Dzb\rhop) = A(\Dm \rhop) + \sqrt{2}A(\Dzb\rhoz) \, .
\end{equation}

This relation can be used to determine $\cos\delta_{D\rho}$, where
$\delta_{D\rho}$ is the phase between the $A_{3/2}$ and $A_{1/2}$ amplitudes,
and $R_{D\rho} = \left| A_{1/2} / \sqrt{2} A_{3/2} \right|$.
In QCD factorization, both of these are expected to be unity up to corrections
due to final state interactions of ${\cal O}(\Lambda_{\rm QCD}/m_Q)$,
where $\Lambda_{\rm QCD}$ is the QCD scale and $m_Q$ is either $m_c$ or
$m_b$~\cite{Neubert:2001sj}.
We obtain constraints on these parameters using the same approach previously
used in the $D^{(*)}\pi$ system~\cite{Aubert:2003sw,Prudent:2008zz}.
Using our result for ${\cal B}(\Bz\to\Dzb\rhoz)$, together with world average
values of ${\cal B}(\Bz\to\Dm\rhop)$, ${\cal B}(\Bp\to\Dzb\rhop)$ and the
ratio of lifetimes $\tau(\Bp)/\tau(\Bz)$~\cite{Amsler:2008zz}, we find
\begin{eqnarray*}
  \cos\delta_{D\rho} & = & 0.998^{+0.133}_{-0.062} \, , \\
  R_{D\rho} & = & 0.68^{+0.15}_{-0.16} \, ,
\end{eqnarray*}
where all sources of uncertainty are combined.
These results suggest the presence of non-factorizable final state interaction
effects that, in contrast to the $D^{(*)}\pi$ system, do not introduce a
significant non-zero phase difference between the isospin amplitudes.

\section{Summary}

We have performed a Dalitz-plot analysis of $\Bz\to\Dzb\pipi$ decays using the whole \babar\ dataset of \bbpairs\ \BB\ events. 
We measure the inclusive branching fraction
$$
{\cal B}(\BztoDzbpipi) = (8.81 \pm 0.18 \pm 0.76 \pm 0.78 \pm 0.11) \times 10^{-4}
$$
where the first uncertainty is statistical, the second is systematic, the third is due to the Dalitz-plot model, and the fourth is due to secondary branching fractions.
We find the Dalitz plot to be composed of contributions from \DstarIIm,
\DstarIm, \rhoIz\ and \fII\ as well as a \pipi\ S-wave, a $D\pi$
nonresonant S-wave term and a virtual \Dstarvm\ contribution. 
We determine their branching fractions:
\begin{eqnarray*}
  {\cal B}(\BztoDstarIImpip)\times{\cal B}(\DstarIIm\to\Dzb\pim) & = &
  (1.80 \pm 0.09 \pm 0.19 \pm 0.37 \pm 0.02) \times 10^{-4} \, , \\
  {\cal B}(\BztoDstarImpip)\times{\cal B}(\DstarIm\to\Dzb\pim) & = &
  (2.18 \pm 0.23 \pm 0.33 \pm 1.15 \pm 0.03) \times 10^{-4} \, , \\
  {\cal B}(\Bz\to\rhoIzDzb) & = &
  (2.98 \pm 0.19 \pm 0.53 \pm 0.93 \pm 0.04) \times 10^{-4} \, , \\
  {\cal B}(\Bz\to\fIIDzb) & = &
  (1.02 \pm 0.12 \pm 0.18 \pm 0.36 \pm 0.03) \times 10^{-4}\, .
\end{eqnarray*}

Our Dalitz plot model differs from that obtained in a previous study of
$\Bz\to\Dzb\pipi$ by Belle~\cite{Kuzmin:2006mw} in that (i) we use the
K-matrix description of the \pipi\ S-wave, instead of including separate
contributions from the $f_0(600)$ ($\sigma$), $f_0(980)$ and $f_0(1370)$
scalar resonances; (ii) we include an additional $D\pi$ nonresonant S-wave
term.
Our results for the inclusive branching fraction and for the color-suppressed
decays $\Bz\to\rhoIzDzb$ and $\Bz\to\fIIDzb$ are consistent with those from
Belle and (for $\rhoIzDzb$) with theoretical
predictions~\cite{Chua:2001br,Keum:2003js}.
However, we find the product branching fractions for the broad and narrow
$D^{**}$ states (\DstarI\ and \DstarII, respectively) to have similar values. 
This result disagrees with the analysis by Belle, which
found a much smaller value for the \DstarI\ branching fraction.

We are grateful for the 
extraordinary contributions of our \pep2\ colleagues in
achieving the excellent luminosity and machine conditions
that have made this work possible.
The success of this project also relies critically on the 
expertise and dedication of the computing organizations that 
support \babar.
The collaborating institutions wish to thank 
SLAC for its support and the kind hospitality extended to them. 
This work is supported by the
US Department of Energy
and National Science Foundation, the
Natural Sciences and Engineering Research Council (Canada),
the Commissariat \`a l'Energie Atomique and
Institut National de Physique Nucl\'eaire et de Physique des Particules
(France), the
Bundesministerium f\"ur Bildung und Forschung and
Deutsche Forschungsgemeinschaft
(Germany), the
Istituto Nazionale di Fisica Nucleare (Italy),
the Foundation for Fundamental Research on Matter (The Netherlands),
the Research Council of Norway, the
Ministry of Education and Science of the Russian Federation, 
Ministerio de Ciencia e Innovaci\'on (Spain), and the
Science and Technology Facilities Council (United Kingdom).
Individuals have received support from 
the Marie-Curie IEF program (European Union), the A. P. Sloan Foundation (USA) 
and the Binational Science Foundation (USA-Israel).

\bibliography{references}

\begin{thebibliography}{55}
\expandafter\ifx\csname natexlab\endcsname\relax\def\natexlab#1{#1}\fi
\expandafter\ifx\csname bibnamefont\endcsname\relax
  \def\bibnamefont#1{#1}\fi
\expandafter\ifx\csname bibfnamefont\endcsname\relax
  \def\bibfnamefont#1{#1}\fi
\expandafter\ifx\csname citenamefont\endcsname\relax
  \def\citenamefont#1{#1}\fi
\expandafter\ifx\csname url\endcsname\relax
  \def\url#1{\texttt{#1}}\fi
\expandafter\ifx\csname urlprefix\endcsname\relax\def\urlprefix{URL }\fi
\providecommand{\bibinfo}[2]{#2}
\providecommand{\eprint}[2][]{\url{#2}}

\bibitem[{\citenamefont{Dalitz}(1953)}]{Dalitz:1953cp}
\bibinfo{author}{\bibfnamefont{R.~H.} \bibnamefont{Dalitz}},
  \bibinfo{journal}{Phil. Mag.} \textbf{\bibinfo{volume}{44}},
  \bibinfo{pages}{1068} (\bibinfo{year}{1953}).

\bibitem[{\citenamefont{Jugeau et~al.}(2005)\citenamefont{Jugeau, Le~Yaouanc,
  Oliver, and Raynal}}]{Jugeau:2005yr}
\bibinfo{author}{\bibfnamefont{F.}~\bibnamefont{Jugeau}},
  \bibinfo{author}{\bibfnamefont{A.}~\bibnamefont{Le~Yaouanc}},
  \bibinfo{author}{\bibfnamefont{L.}~\bibnamefont{Oliver}}, \bibnamefont{and}
  \bibinfo{author}{\bibfnamefont{J.~C.} \bibnamefont{Raynal}},
  \bibinfo{journal}{Phys. Rev.} \textbf{\bibinfo{volume}{D72}},
  \bibinfo{pages}{094010} (\bibinfo{year}{2005}), \eprint{hep-ph/0504206}.

\bibitem[{\citenamefont{Liventsev et~al.}(2008)}]{:2007rb}
\bibinfo{author}{\bibfnamefont{D.}~\bibnamefont{Liventsev}}
  \bibnamefont{et~al.} (\bibinfo{collaboration}{Belle Collaboration}),
  \bibinfo{journal}{Phys. Rev.} \textbf{\bibinfo{volume}{D77}},
  \bibinfo{pages}{091503} (\bibinfo{year}{2008}), \eprint{arXiv:0711.3252
  [hep-ex]}.

\bibitem[{\citenamefont{Aubert et~al.}(2008{\natexlab{a}})}]{Aubert:2008ea}
\bibinfo{author}{\bibfnamefont{B.}~\bibnamefont{Aubert}} \bibnamefont{et~al.}
  (\bibinfo{collaboration}{\babar\ Collaboration}), \bibinfo{journal}{Phys.
  Rev. Lett.} \textbf{\bibinfo{volume}{101}}, \bibinfo{pages}{261802}
  (\bibinfo{year}{2008}{\natexlab{a}}), \eprint{arXiv:0808.0528 [hep-ex]}.

\bibitem[{\citenamefont{Morenas et~al.}(1997)\citenamefont{Morenas, Le~Yaouanc,
  Oliver, Pene, and Raynal}}]{Morenas:1997nk}
\bibinfo{author}{\bibfnamefont{V.}~\bibnamefont{Morenas}},
  \bibinfo{author}{\bibfnamefont{A.}~\bibnamefont{Le~Yaouanc}},
  \bibinfo{author}{\bibfnamefont{L.}~\bibnamefont{Oliver}},
  \bibinfo{author}{\bibfnamefont{O.}~\bibnamefont{Pene}}, \bibnamefont{and}
  \bibinfo{author}{\bibfnamefont{J.~C.} \bibnamefont{Raynal}},
  \bibinfo{journal}{Phys. Rev.} \textbf{\bibinfo{volume}{D56}},
  \bibinfo{pages}{5668} (\bibinfo{year}{1997}), \eprint{hep-ph/9706265}.

\bibitem[{\citenamefont{Dai and Huang}(1999)}]{Dai:1998ca}
\bibinfo{author}{\bibfnamefont{Y.-b.} \bibnamefont{Dai}} \bibnamefont{and}
  \bibinfo{author}{\bibfnamefont{M.-q.} \bibnamefont{Huang}},
  \bibinfo{journal}{Phys. Rev.} \textbf{\bibinfo{volume}{D59}},
  \bibinfo{pages}{034018} (\bibinfo{year}{1999}), \eprint{hep-ph/9807461}.

\bibitem[{\citenamefont{Uraltsev}(2001)}]{Uraltsev:2000ce}
\bibinfo{author}{\bibfnamefont{N.}~\bibnamefont{Uraltsev}},
  \bibinfo{journal}{Phys. Lett.} \textbf{\bibinfo{volume}{B501}},
  \bibinfo{pages}{86} (\bibinfo{year}{2001}), \eprint{hep-ph/0011124}.

\bibitem[{\citenamefont{Le~Yaouanc et~al.}(2001)\citenamefont{Le~Yaouanc,
  Oliver, Pene, Raynal, and Morenas}}]{LeYaouanc:2001nk}
\bibinfo{author}{\bibfnamefont{A.}~\bibnamefont{Le~Yaouanc}},
  \bibinfo{author}{\bibfnamefont{L.}~\bibnamefont{Oliver}},
  \bibinfo{author}{\bibfnamefont{O.}~\bibnamefont{Pene}},
  \bibinfo{author}{\bibfnamefont{J.~C.} \bibnamefont{Raynal}},
  \bibnamefont{and} \bibinfo{author}{\bibfnamefont{V.}~\bibnamefont{Morenas}},
  \bibinfo{journal}{Phys. Lett.} \textbf{\bibinfo{volume}{B520}},
  \bibinfo{pages}{25} (\bibinfo{year}{2001}), \eprint{hep-ph/0105247}.

\bibitem[{\citenamefont{Blossier et~al.}(2009)\citenamefont{Blossier, Wagner,
  and Pene}}]{Blossier:2009vy}
\bibinfo{author}{\bibfnamefont{B.}~\bibnamefont{Blossier}},
  \bibinfo{author}{\bibfnamefont{M.}~\bibnamefont{Wagner}}, \bibnamefont{and}
  \bibinfo{author}{\bibfnamefont{O.}~\bibnamefont{Pene}}
  (\bibinfo{collaboration}{European Twisted Mass Collaboration}),
  \bibinfo{journal}{JHEP} \textbf{\bibinfo{volume}{06}}, \bibinfo{pages}{022}
  (\bibinfo{year}{2009}), \eprint{arXiv:0903.2298 [hep-lat]}.

\bibitem[{\citenamefont{Bauer et~al.}(1987)\citenamefont{Bauer, Stech, and
  Wirbel}}]{Bauer:1986bm}
\bibinfo{author}{\bibfnamefont{M.}~\bibnamefont{Bauer}},
  \bibinfo{author}{\bibfnamefont{B.}~\bibnamefont{Stech}}, \bibnamefont{and}
  \bibinfo{author}{\bibfnamefont{M.}~\bibnamefont{Wirbel}},
  \bibinfo{journal}{Z. Phys.} \textbf{\bibinfo{volume}{C34}},
  \bibinfo{pages}{103} (\bibinfo{year}{1987}).

\bibitem[{\citenamefont{Neubert and Petrov}(2001)}]{Neubert:2001sj}
\bibinfo{author}{\bibfnamefont{M.}~\bibnamefont{Neubert}} \bibnamefont{and}
  \bibinfo{author}{\bibfnamefont{A.~A.} \bibnamefont{Petrov}},
  \bibinfo{journal}{Phys. Lett.} \textbf{\bibinfo{volume}{B519}},
  \bibinfo{pages}{50} (\bibinfo{year}{2001}), \eprint{hep-ph/0108103}.

\bibitem[{\citenamefont{Chua et~al.}(2002)\citenamefont{Chua, Hou, and
  Yang}}]{Chua:2001br}
\bibinfo{author}{\bibfnamefont{C.-K.} \bibnamefont{Chua}},
  \bibinfo{author}{\bibfnamefont{W.-S.} \bibnamefont{Hou}}, \bibnamefont{and}
  \bibinfo{author}{\bibfnamefont{K.-C.} \bibnamefont{Yang}},
  \bibinfo{journal}{Phys. Rev.} \textbf{\bibinfo{volume}{D65}},
  \bibinfo{pages}{096007} (\bibinfo{year}{2002}), \eprint{hep-ph/0112148}.

\bibitem[{\citenamefont{Mantry et~al.}(2003)\citenamefont{Mantry, Pirjol, and
  Stewart}}]{Mantry:2003uz}
\bibinfo{author}{\bibfnamefont{S.}~\bibnamefont{Mantry}},
  \bibinfo{author}{\bibfnamefont{D.}~\bibnamefont{Pirjol}}, \bibnamefont{and}
  \bibinfo{author}{\bibfnamefont{I.~W.} \bibnamefont{Stewart}},
  \bibinfo{journal}{Phys. Rev.} \textbf{\bibinfo{volume}{D68}},
  \bibinfo{pages}{114009} (\bibinfo{year}{2003}), \eprint{hep-ph/0306254}.

\bibitem[{\citenamefont{Rosner}(1999)}]{Rosner:1999zm}
\bibinfo{author}{\bibfnamefont{J.~L.} \bibnamefont{Rosner}},
  \bibinfo{journal}{Phys. Rev.} \textbf{\bibinfo{volume}{D60}},
  \bibinfo{pages}{074029} (\bibinfo{year}{1999}), \eprint{hep-ph/9903543}.

\bibitem[{\citenamefont{Cabibbo}(1963)}]{Cabibbo:1963yz}
\bibinfo{author}{\bibfnamefont{N.}~\bibnamefont{Cabibbo}},
  \bibinfo{journal}{Phys. Rev. Lett.} \textbf{\bibinfo{volume}{10}},
  \bibinfo{pages}{531} (\bibinfo{year}{1963}).

\bibitem[{\citenamefont{Kobayashi and Maskawa}(1973)}]{Kobayashi:1973fv}
\bibinfo{author}{\bibfnamefont{M.}~\bibnamefont{Kobayashi}} \bibnamefont{and}
  \bibinfo{author}{\bibfnamefont{T.}~\bibnamefont{Maskawa}},
  \bibinfo{journal}{Prog. Theor. Phys.} \textbf{\bibinfo{volume}{49}},
  \bibinfo{pages}{652} (\bibinfo{year}{1973}).

\bibitem[{\citenamefont{Fleischer}(2003{\natexlab{a}})}]{Fleischer:2003ai}
\bibinfo{author}{\bibfnamefont{R.}~\bibnamefont{Fleischer}},
  \bibinfo{journal}{Phys. Lett.} \textbf{\bibinfo{volume}{B562}},
  \bibinfo{pages}{234} (\bibinfo{year}{2003}{\natexlab{a}}),
  \eprint{hep-ph/0301255}.

\bibitem[{\citenamefont{Fleischer}(2003{\natexlab{b}})}]{Fleischer:2003aj}
\bibinfo{author}{\bibfnamefont{R.}~\bibnamefont{Fleischer}},
  \bibinfo{journal}{Nucl. Phys.} \textbf{\bibinfo{volume}{B659}},
  \bibinfo{pages}{321} (\bibinfo{year}{2003}{\natexlab{b}}),
  \eprint{hep-ph/0301256}.

\bibitem[{\citenamefont{Grossman and Worah}(1997)}]{Grossman:1996ke}
\bibinfo{author}{\bibfnamefont{Y.}~\bibnamefont{Grossman}} \bibnamefont{and}
  \bibinfo{author}{\bibfnamefont{M.~P.} \bibnamefont{Worah}},
  \bibinfo{journal}{Phys. Lett.} \textbf{\bibinfo{volume}{B395}},
  \bibinfo{pages}{241} (\bibinfo{year}{1997}), \eprint{hep-ph/9612269}.

\bibitem[{\citenamefont{Charles et~al.}(1998)\citenamefont{Charles, Le~Yaouanc,
  Oliver, Pene, and Raynal}}]{Charles:1998vf}
\bibinfo{author}{\bibfnamefont{J.}~\bibnamefont{Charles}},
  \bibinfo{author}{\bibfnamefont{A.}~\bibnamefont{Le~Yaouanc}},
  \bibinfo{author}{\bibfnamefont{L.}~\bibnamefont{Oliver}},
  \bibinfo{author}{\bibfnamefont{O.}~\bibnamefont{Pene}}, \bibnamefont{and}
  \bibinfo{author}{\bibfnamefont{J.~C.} \bibnamefont{Raynal}},
  \bibinfo{journal}{Phys. Lett.} \textbf{\bibinfo{volume}{B425}},
  \bibinfo{pages}{375} (\bibinfo{year}{1998}), \eprint{hep-ph/9801363}.

\bibitem[{\citenamefont{Latham and Gershon}(2009)}]{Latham:2008zs}
\bibinfo{author}{\bibfnamefont{T.}~\bibnamefont{Latham}} \bibnamefont{and}
  \bibinfo{author}{\bibfnamefont{T.}~\bibnamefont{Gershon}},
  \bibinfo{journal}{J. Phys.} \textbf{\bibinfo{volume}{G36}},
  \bibinfo{pages}{025006} (\bibinfo{year}{2009}), \eprint{arXiv:0809.0872
  [hep-ph]}.

\bibitem[{\citenamefont{Kuzmin et~al.}(2007)}]{Kuzmin:2006mw}
\bibinfo{author}{\bibfnamefont{A.}~\bibnamefont{Kuzmin}} \bibnamefont{et~al.}
  (\bibinfo{collaboration}{Belle Collaboration}), \bibinfo{journal}{Phys. Rev.}
  \textbf{\bibinfo{volume}{D76}}, \bibinfo{pages}{012006}
  (\bibinfo{year}{2007}), \eprint{hep-ex/0611054}.

\bibitem[{\citenamefont{Aubert et~al.}(2009{\natexlab{a}})}]{Aubert:2009wg}
\bibinfo{author}{\bibfnamefont{B.}~\bibnamefont{Aubert}} \bibnamefont{et~al.}
  (\bibinfo{collaboration}{\babar\ Collaboration}), \bibinfo{journal}{Phys.
  Rev.} \textbf{\bibinfo{volume}{D79}}, \bibinfo{pages}{112004}
  (\bibinfo{year}{2009}{\natexlab{a}}), \eprint{arXiv:0901.1291 [hep-ex]}.

\bibitem[{\citenamefont{Abe et~al.}(2004)}]{Abe:2003zm}
\bibinfo{author}{\bibfnamefont{K.}~\bibnamefont{Abe}} \bibnamefont{et~al.}
  (\bibinfo{collaboration}{Belle Collaboration}), \bibinfo{journal}{Phys. Rev.}
  \textbf{\bibinfo{volume}{D69}}, \bibinfo{pages}{112002}
  (\bibinfo{year}{2004}), \eprint{hep-ex/0307021}.

\bibitem[{\citenamefont{Aubert et~al.}(2002)}]{Aubert:2001tu}
\bibinfo{author}{\bibfnamefont{B.}~\bibnamefont{Aubert}} \bibnamefont{et~al.}
  (\bibinfo{collaboration}{\babar\ Collaboration}), \bibinfo{journal}{Nucl.
  Instrum. Methods Phys. Res., Sect. A} \textbf{\bibinfo{volume}{479}},
  \bibinfo{pages}{1} (\bibinfo{year}{2002}).

\bibitem[{\citenamefont{Amsler et~al.}(2008)}]{Amsler:2008zz}
\bibinfo{author}{\bibfnamefont{C.}~\bibnamefont{Amsler}} \bibnamefont{et~al.}
  (\bibinfo{collaboration}{Particle Data Group}), \bibinfo{journal}{Phys.
  Lett.} \textbf{\bibinfo{volume}{B667}}, \bibinfo{pages}{1}
  (\bibinfo{year}{2008}).

\bibitem[{\citenamefont{Albrecht et~al.}(1990)}]{Albrecht:1990cs}
\bibinfo{author}{\bibfnamefont{H.}~\bibnamefont{Albrecht}} \bibnamefont{et~al.}
  (\bibinfo{collaboration}{ARGUS Collaboration}), \bibinfo{journal}{Z. Phys.}
  \textbf{\bibinfo{volume}{C48}}, \bibinfo{pages}{543} (\bibinfo{year}{1990}).

\bibitem[{\citenamefont{Fleming}(1964)}]{Fleming:1964zz}
\bibinfo{author}{\bibfnamefont{G.~N.} \bibnamefont{Fleming}},
  \bibinfo{journal}{Phys. Rev.} \textbf{\bibinfo{volume}{135}},
  \bibinfo{pages}{B551} (\bibinfo{year}{1964}).

\bibitem[{\citenamefont{Morgan}(1968)}]{Morgan:1968zz}
\bibinfo{author}{\bibfnamefont{D.}~\bibnamefont{Morgan}},
  \bibinfo{journal}{Phys. Rev.} \textbf{\bibinfo{volume}{166}},
  \bibinfo{pages}{1731} (\bibinfo{year}{1968}).

\bibitem[{\citenamefont{Herndon et~al.}(1975)\citenamefont{Herndon, Soding, and
  Cashmore}}]{Herndon:1973yn}
\bibinfo{author}{\bibfnamefont{D.}~\bibnamefont{Herndon}},
  \bibinfo{author}{\bibfnamefont{P.}~\bibnamefont{Soding}}, \bibnamefont{and}
  \bibinfo{author}{\bibfnamefont{R.~J.} \bibnamefont{Cashmore}},
  \bibinfo{journal}{Phys. Rev.} \textbf{\bibinfo{volume}{D11}},
  \bibinfo{pages}{3165} (\bibinfo{year}{1975}).

\bibitem[{\citenamefont{Garmash et~al.}(2005)}]{Garmash:2004wa}
\bibinfo{author}{\bibfnamefont{A.}~\bibnamefont{Garmash}} \bibnamefont{et~al.}
  (\bibinfo{collaboration}{Belle Collaboration}), \bibinfo{journal}{Phys. Rev.}
  \textbf{\bibinfo{volume}{D71}}, \bibinfo{pages}{092003}
  (\bibinfo{year}{2005}).

\bibitem[{\citenamefont{Chung et~al.}(1995)}]{Chung:1995dx}
\bibinfo{author}{\bibfnamefont{S.~U.} \bibnamefont{Chung}}
  \bibnamefont{et~al.}, \bibinfo{journal}{Annalen Phys.}
  \textbf{\bibinfo{volume}{4}}, \bibinfo{pages}{404} (\bibinfo{year}{1995}).

\bibitem[{\citenamefont{Aitchison}(1972)}]{Aitchison:1972ay}
\bibinfo{author}{\bibfnamefont{I.~J.~R.} \bibnamefont{Aitchison}},
  \bibinfo{journal}{Nucl. Phys.} \textbf{\bibinfo{volume}{A189}},
  \bibinfo{pages}{417} (\bibinfo{year}{1972}).

\bibitem[{\citenamefont{Anisovich and Sarantsev}(2003)}]{Anisovich:2002ij}
\bibinfo{author}{\bibfnamefont{V.~V.} \bibnamefont{Anisovich}}
  \bibnamefont{and} \bibinfo{author}{\bibfnamefont{A.~V.}
  \bibnamefont{Sarantsev}}, \bibinfo{journal}{Eur. Phys. J.}
  \textbf{\bibinfo{volume}{A16}}, \bibinfo{pages}{229} (\bibinfo{year}{2003}),
  \eprint{hep-ph/0204328}.

\bibitem[{\citenamefont{Blatt and Weisskopf}(1952)}]{blatt-weisskopf}
\bibinfo{author}{\bibfnamefont{J.}~\bibnamefont{Blatt}} \bibnamefont{and}
  \bibinfo{author}{\bibfnamefont{V.~E.} \bibnamefont{Weisskopf}},
  \emph{\bibinfo{title}{Theoretical Nuclear Physics}} (\bibinfo{publisher}{J.
  Wiley (New York)}, \bibinfo{year}{1952}).

\bibitem[{\citenamefont{Zemach}(1964)}]{Zemach:1963bc}
\bibinfo{author}{\bibfnamefont{C.}~\bibnamefont{Zemach}},
  \bibinfo{journal}{Phys. Rev.} \textbf{\bibinfo{volume}{133}},
  \bibinfo{pages}{B1201} (\bibinfo{year}{1964}).

\bibitem[{\citenamefont{Zemach}(1965)}]{Zemach:1968zz}
\bibinfo{author}{\bibfnamefont{C.}~\bibnamefont{Zemach}},
  \bibinfo{journal}{Phys. Rev.} \textbf{\bibinfo{volume}{140}},
  \bibinfo{pages}{B97} (\bibinfo{year}{1965}).

\bibitem[{\citenamefont{Aubert et~al.}(2005{\natexlab{a}})}]{Aubert:2005sk}
\bibinfo{author}{\bibfnamefont{B.}~\bibnamefont{Aubert}} \bibnamefont{et~al.}
  (\bibinfo{collaboration}{\babar\ Collaboration}), \bibinfo{journal}{Phys.
  Rev.} \textbf{\bibinfo{volume}{D72}}, \bibinfo{pages}{052002}
  (\bibinfo{year}{2005}{\natexlab{a}}), \eprint{hep-ex/0507025}.

\bibitem[{\citenamefont{Aubert et~al.}(2005{\natexlab{b}})}]{Aubert:2005ce}
\bibinfo{author}{\bibfnamefont{B.}~\bibnamefont{Aubert}} \bibnamefont{et~al.}
  (\bibinfo{collaboration}{\babar\ Collaboration}), \bibinfo{journal}{Phys.
  Rev.} \textbf{\bibinfo{volume}{D72}}, \bibinfo{pages}{072003}
  (\bibinfo{year}{2005}{\natexlab{b}}), \bibinfo{note}{[Erratum-ibid.\ D{\bf
  74} 099903 (2006)]}, \eprint{hep-ex/0507004}.

\bibitem[{\citenamefont{Aubert et~al.}(2008{\natexlab{b}})}]{Aubert:2008bj}
\bibinfo{author}{\bibfnamefont{B.}~\bibnamefont{Aubert}} \bibnamefont{et~al.}
  (\bibinfo{collaboration}{\babar\ Collaboration}), \bibinfo{journal}{Phys.
  Rev.} \textbf{\bibinfo{volume}{D78}}, \bibinfo{pages}{012004}
  (\bibinfo{year}{2008}{\natexlab{b}}), \eprint{arXiv:0803.4451 [hep-ex]}.

\bibitem[{\citenamefont{Aubert et~al.}(2009{\natexlab{b}})}]{:2009az}
\bibinfo{author}{\bibfnamefont{B.}~\bibnamefont{Aubert}} \bibnamefont{et~al.}
  (\bibinfo{collaboration}{\babar\ Collaboration}), \bibinfo{journal}{Phys.
  Rev.} \textbf{\bibinfo{volume}{D79}}, \bibinfo{pages}{072006}
  (\bibinfo{year}{2009}{\natexlab{b}}), \eprint{arXiv:0902.2051 [hep-ex]}.

\bibitem[{\citenamefont{Pivk and Le~Diberder}(2005)}]{Pivk:2004ty}
\bibinfo{author}{\bibfnamefont{M.}~\bibnamefont{Pivk}} \bibnamefont{and}
  \bibinfo{author}{\bibfnamefont{F.~R.} \bibnamefont{Le~Diberder}},
  \bibinfo{journal}{Nucl. Instrum. Meth.} \textbf{\bibinfo{volume}{A555}},
  \bibinfo{pages}{356} (\bibinfo{year}{2005}), \eprint{physics/0402083}.

\bibitem[{\citenamefont{Gounaris and Sakurai}(1968)}]{Gounaris:1968mw}
\bibinfo{author}{\bibfnamefont{G.~J.} \bibnamefont{Gounaris}} \bibnamefont{and}
  \bibinfo{author}{\bibfnamefont{J.~J.} \bibnamefont{Sakurai}},
  \bibinfo{journal}{Phys. Rev. Lett.} \textbf{\bibinfo{volume}{21}},
  \bibinfo{pages}{244} (\bibinfo{year}{1968}).

\bibitem[{\citenamefont{Bugg}(2003)}]{Bugg:2003kj}
\bibinfo{author}{\bibfnamefont{D.~V.} \bibnamefont{Bugg}},
  \bibinfo{journal}{Phys. Lett.} \textbf{\bibinfo{volume}{B572}},
  \bibinfo{pages}{1} (\bibinfo{year}{2003}).

\bibitem[{\citenamefont{Flatte}(1976)}]{Flatte:1976xu}
\bibinfo{author}{\bibfnamefont{S.~M.} \bibnamefont{Flatte}},
  \bibinfo{journal}{Phys. Lett.} \textbf{\bibinfo{volume}{B63}},
  \bibinfo{pages}{224} (\bibinfo{year}{1976}).

\bibitem[{\citenamefont{Bugg}(2009)}]{Bugg:2009tu}
\bibinfo{author}{\bibfnamefont{D.~V.} \bibnamefont{Bugg}}, \bibinfo{journal}{J.
  Phys.} \textbf{\bibinfo{volume}{G36}}, \bibinfo{pages}{075003}
  (\bibinfo{year}{2009}), \eprint{arXiv:0901.2217 [hep-ph]}.

\bibitem[{\citenamefont{Aubert et~al.}()}]{Aubert:201zzz}
\bibinfo{author}{\bibfnamefont{B.}~\bibnamefont{Aubert}} \bibnamefont{et~al.}
  (\bibinfo{collaboration}{\babar\ Collaboration}), \bibinfo{note}{in
  preparation}.

\bibitem[{\citenamefont{Chiang and Rosner}(2003)}]{Chiang:2002tv}
\bibinfo{author}{\bibfnamefont{C.-W.} \bibnamefont{Chiang}} \bibnamefont{and}
  \bibinfo{author}{\bibfnamefont{J.~L.} \bibnamefont{Rosner}},
  \bibinfo{journal}{Phys. Rev.} \textbf{\bibinfo{volume}{D67}},
  \bibinfo{pages}{074013} (\bibinfo{year}{2003}), \eprint{hep-ph/0212274}.

\bibitem[{\citenamefont{Mantry}(2004)}]{Mantry:2004pg}
\bibinfo{author}{\bibfnamefont{S.}~\bibnamefont{Mantry}},
  \bibinfo{journal}{Phys. Rev.} \textbf{\bibinfo{volume}{D70}},
  \bibinfo{pages}{114006} (\bibinfo{year}{2004}), \eprint{hep-ph/0405290}.

\bibitem[{\citenamefont{Aubert et~al.}(2004)}]{Aubert:2003sw}
\bibinfo{author}{\bibfnamefont{B.}~\bibnamefont{Aubert}} \bibnamefont{et~al.}
  (\bibinfo{collaboration}{\babar\ Collaboration}), \bibinfo{journal}{Phys.
  Rev.} \textbf{\bibinfo{volume}{D69}}, \bibinfo{pages}{032004}
  (\bibinfo{year}{2004}), \eprint{hep-ex/0310028}.

\bibitem[{\citenamefont{Prudent}(2008)}]{Prudent:2008zz}
\bibinfo{author}{\bibfnamefont{X.}~\bibnamefont{Prudent}}
  (\bibinfo{year}{2008}), \bibinfo{note}{lAPP-T-2008-01}.

\bibitem[{\citenamefont{Keum et~al.}(2004)\citenamefont{Keum, Kurimoto, Li, Lu,
  and Sanda}}]{Keum:2003js}
\bibinfo{author}{\bibfnamefont{Y.-Y.} \bibnamefont{Keum}},
  \bibinfo{author}{\bibfnamefont{T.}~\bibnamefont{Kurimoto}},
  \bibinfo{author}{\bibfnamefont{H.~N.} \bibnamefont{Li}},
  \bibinfo{author}{\bibfnamefont{C.-D.} \bibnamefont{Lu}}, \bibnamefont{and}
  \bibinfo{author}{\bibfnamefont{A.~I.} \bibnamefont{Sanda}},
  \bibinfo{journal}{Phys. Rev.} \textbf{\bibinfo{volume}{D69}},
  \bibinfo{pages}{094018} (\bibinfo{year}{2004}), \eprint{hep-ph/0305335}.

\bibitem[{\citenamefont{Aston et~al.}(1988)}]{lass}
\bibinfo{author}{\bibfnamefont{D.}~\bibnamefont{Aston}} \bibnamefont{et~al.}
  (\bibinfo{collaboration}{LASS Collaboration}), \bibinfo{journal}{Nucl. Phys.}
  \textbf{\bibinfo{volume}{B296}}, \bibinfo{pages}{493} (\bibinfo{year}{1988}).

\bibitem[{\citenamefont{Asner}(2008)}]{Asner:2008zzb}
\bibinfo{author}{\bibfnamefont{D.}~\bibnamefont{Asner}},
  \bibinfo{journal}{Phys. Lett.} \textbf{\bibinfo{volume}{B667}},
  \bibinfo{pages}{774} (\bibinfo{year}{2008}).

\bibitem[{\citenamefont{Aubert et~al.}(2008{\natexlab{c}})}]{Aubert:2008bd}
\bibinfo{author}{\bibfnamefont{B.}~\bibnamefont{Aubert}} \bibnamefont{et~al.}
  (\bibinfo{collaboration}{\babar\ Collaboration}), \bibinfo{journal}{Phys.
  Rev.} \textbf{\bibinfo{volume}{D78}}, \bibinfo{pages}{034023}
  (\bibinfo{year}{2008}{\natexlab{c}}), \eprint{arXiv:0804.2089 [hep-ex]}.

\end{thebibliography}
\bibliographystyle{apsrev}

\appendix*
\subsubsection{Appendix: K-matrix description of \pipi\ S wave}
\label{subsubsec:KMatrix}

The K-matrix formalism gives a physical description of broad overlapping
states -- {\it i.e.} it does not violate unitarity, unlike the more
conventional ``sum of Breit--Wigners'' approach.
The K-matrix formalism can be shown to reduce to more familiar forms (the
Breit--Wigner lineshape for single resonances, the Flatt\'e
lineshape~\cite{Flatte:1976xu} for coupled channels, the LASS
formula~\cite{lass} for broad resonances interfering with nonresonant terms).
Detailed descriptions of the K-matrix formalism can be found in various
references~\cite{Chung:1995dx,Aitchison:1972ay,Anisovich:2002ij,Asner:2008zzb}.
Here we give an outline of the salient features. 

The scattering (``S'') matrix describes transitions from initial states
$|i\rangle$ into final states $|f\rangle$ ($S_{if}=\langle f|S|i\rangle$), and
can be written
\begin{equation}
  \label{eq:Smatrix_def}
  S = I+2i \{ \rho^\dag \}^{1/2} T \{\rho\}^{1/2} \, , 
\end{equation}
where $I$ is the identity matrix, 
$\rho$ is a diagonal phase-space matrix, with
elements $\rho_{ii}=2q_i/m$ with $q_i$ the threshold momentum, 
and $T$ is the transition matrix.
The unitarity requirement ($SS^\dag=S^\dag S=I$) gives
\begin{equation}
  \label{eq:kmatrixT}
  \left(T^{-1}+i\rho\right)^\dag = T^{-1}+i\rho\,, 
  \qquad\Rightarrow\qquad K^{-1}=T^{-1}+i\rho\,,
\end{equation}
where $K$ is a Lorentz-invariant and Hermitian matrix which describes the
decay process.
This formalism was developed for scattering processes, but can also be applied
to Dalitz plot analyses, with the assumption that the two ``scattering''
products do not interact with the third bachelor
particle~\cite{Aitchison:1972ay}.
However, it is also necessary to include a process-dependent production
vector, which accounts for the relative production rates of the different
states $|i\rangle$. 
We refer to the ``K-matrix amplitude'' as a product of the production vector
$P$ and the (matrix) propagator $\left(I-i K\rho\right)^{-1}$:
\begin{equation}
  A_i = \left(I-i K\rho\right)^{-1}_{ij} P_j
\end{equation}

The K matrix is expressed as
\begin{equation}
  K_{ij}(s) = 
  \left[
    f_{ij}^{\rm scatt}\frac{1-s_0^{\rm scatt}}{s-s_0^{\rm scatt}} + 
    \sum_\alpha \frac{g_i^{(\alpha)}g_j^{(\alpha)}}{m_\alpha^2-s}
  \right]
  \left\{
    \frac{1-s_{A0}}{s-s_{A0}}\left(s-\frac{s_Am_\pi^2}{2}\right)
  \right\}\,,
\end{equation}
where the factor $g_i^{(\alpha)}$ is the real coupling constant of the K
matrix pole $\alpha$ (with mass $m_\alpha$) to meson channel $i$, 
the parameters $f_{ij}^{\rm scatt}$ and $s_0^{\rm scatt}$ describe a smooth
part for the K-matrix elements, 
and the last factor accounts for the so-called ``Adler zero'', and suppresses
kinematically fake singularities near \pipi\ production threshold ($s$
represents the square of the \pipi\ invariant mass).
The K-matrix parameters are determined from global fits to scattering data
experiments below 1900\mevcc~\cite{Anisovich:2002ij}.  Note that the
phase-space for \Bz\to\Dzb\pipi\ extends beyond this limit, and that the
K-matrix amplitude in this high-\pipi\ invariant mass region is therefore an
extrapolation.

The parameters unique to the production vector, by contrast, must be
determined from our data.
The $P$ vector is given by
\begin{equation}
  \label{eq:pvector}
  P_j(s) = \left[
    f_{1j}^{\rm prod}\frac{1-s_0^{\rm prod}}{s-s_0^{\rm prod}} + 
    \sum_\alpha\frac{\beta_\alpha g_j^{(\alpha)}}{m_\alpha^2-s}
  \right]\,,
\end{equation}
where as before the first term in the square brackets
is nonresonant-like (``slowly varying''), and the second term is resonant-like.
Hence the free parameters in the Dalitz plot fit are the complex coupling
and production vector parameters $\beta_\alpha$ and $f_{1j}^{\rm prod}$
(we use a fixed value of $s_0^{\rm prod}$). 
The index $j$ runs over the open channels for the $\pi\pi$ $S$-wave, which
are: $\pi\pi$, $KK$, $\eta\eta$, $\eta\eta^\prime$ and $4\pi$ (or
multi-meson).
At higher masses there are in principle more open channels, but this is
not expected to affect the results significantly.
Global fits to the scattering data determine the number of poles and their
parameters.
We use a 5 pole approximation, and give the values of all fixed parameters in
the K-matrix model in \tabref{AS}. 
Note that all $f_{ij}^{\rm prod}=0$ for $i\not =1$ since we are interested
only in the $\pi\pi$ final state.

\begin{table}[hbt!]
  \caption{
    K-matrix parameters from a global analysis of the available $\pi\pi$
    scattering data from threshold up to $1900$~\mevcc~\cite{Anisovich:2002ij,Aubert:2008bd}.
    Masses and coupling constants are given in~\gevcc.
  }
  \label{tab:AS} 
\begin{ruledtabular}
\begin{tabular}{cccccc}
$m_{\alpha}$ & $g_{\pipi}^\alpha$ & $g_{K\Kbar}^\alpha$ & $g_{4\pi}^\alpha$ & $g_{\eta\eta}^\alpha$ & $g_{\eta\eta^{\prime}}^\alpha$  \\
\hline
     $\phm0.65100$  &   $\phm0.22889$  &  $-0.55377$      &    $\phm0.00000$   &  $-0.39899$      &  $-0.34639$ \\
     $\phm1.20360$  &   $\phm0.94128$  &  $\phm0.55095$   &    $\phm0.00000$   &  $\phm0.39065$   &  $\phm0.31503$ \\
     $\phm1.55817$  &   $\phm0.36856$  &  $\phm0.23888$   &    $\phm0.55639$   &  $\phm0.18340$   &  $\phm0.18681$ \\
     $\phm1.21000$  &   $\phm0.33650$  &  $\phm0.40907$   &    $\phm0.85679$   &  $\phm0.19906$   &  $-0.00984$ \\
     $\phm1.82206$  &   $\phm0.18171$  &  $-0.17558$      &    $-0.79658$      &  $-0.00355$      &  $\phm0.22358$ \\
\hline
     &    $f^{\rm scatt}_{11}$  &   $f^{\rm scatt}_{12}$  &   $f^{\rm scatt}_{13}$  &   $f^{\rm scatt}_{14}$  &   $f^{\rm scatt}_{15}$ \\
    &    $\phm0.23399$         &   $\phm0.15044$         &   $-0.20545$            &     $\phm0.32825$       &   $\phm0.35412$   \\	
    \hline
    & $s^{\rm scatt}_0$    & $s_0^{\rm prod}$ & $s_{A0}$      &     $s_{A}$ \\
    & $-3.92637$         & $-3.0$          & $-0.15$       &     $1$ \\
\end{tabular}
\end{ruledtabular}
\end{table}

\end{document}